\documentclass[12pt,preprint]{aastex}

\def\spose#1{\hbox to 0pt{#1\hss}}

\def\kms{\ifmmode {\rm\,km\,s^{-1}}\else ${\rm\,km\,s^{-1}}$\fi}

\def\kmsmpc{\ifmmode {\rm\,km\,s^{-1}\,Mpc^{-1}}\else ${\rm\,km\,s^{-1}\,Mpc^{-1}}$\fi}

\def\ergps{\ifmmode {\rm\,erg\,s^{-1}}\else ${\rm\,erg\,s^{-1}}$\fi}
\def\ergpscm2{\ifmmode {\rm\,erg\,s^{-1}\,cm^{-2}}\else
    ${\rm\,erg\,s^{-1}\,cm^{-2}}$\fi}

\def\deg{\ifmmode {^{\circ}}\else {$^\circ$}\fi}
\def\degr{\ifmmode {^{\circ}}\else {$^\circ$}\fi}
\def\degs{\ifmmode {^{\circ}}\else {$^\circ$}\fi}

\def\h3Mpc{h^{-3}{\rm Mpc}^3}

\def\arcsec{\ifmmode {^{\prime\prime}}\else $^{\prime\prime}$\fi}
\def\asec{\ifmmode {^{\prime\prime}}\else $^{\prime\prime}$\fi}
\def\arcmin{\ifmmode {^{\prime}}\else $^{\prime}$\fi}
\def\amin{\ifmmode {^{\prime}}\else $^{\prime}$\fi}

\def\secper{\ifmmode \rlap.{^{s}}\else $\rlap{.}{^{s}} $\fi}
\def\minper{\ifmmode \rlap.{^{m}}\else $\rlap{.}{^m} $\fi}
\def\secspt{\ifmmode \rlap.{^{\prime\prime}}\else
    $\rlap.{^{\prime\prime}}$\fi}
\def\arcsper{\ifmmode \rlap.{^{\prime\prime}}\else
    $\rlap.{^{\prime\prime}}$\fi}
\def\minspt{\ifmmode \rlap.{^{\prime}}\else
    $\rlap.{^{\prime}}$\fi}
\def\arcmper{\ifmmode \rlap.{^{\prime}}\else
    $\rlap.{^{\prime}}$\fi}
\def\spose#1{\hbox to 0pt{#1\hss}}
\def\simlt{\mathrel{\spose{\lower 3pt\hbox{$\mathchar"218$}}
     \raise 2.0pt\hbox{$\mathchar"13C$}}}
\def\simgt{\mathrel{\spose{\lower 3pt\hbox{$\mathchar"218$}}
     \raise 2.0pt\hbox{$\mathchar"13E$}}}

\def\U300{\ifmmode{U_{300}}\else{$U_{300}$}\fi}
\def\B450{\ifmmode{B_{450}}\else{$B_{450}$}\fi}
\def\V606{\ifmmode{V_{606}}\else{$V_{606}$}\fi}
\def\I814{\ifmmode{I_{814}}\else{$I_{814}$}\fi}
\def\J110{\ifmmode{J_{110}}\else{$J_{110}$}\fi}
\def\H160{\ifmmode{H_{160}}\else{$H_{160}$}\fi}

\begin{document}

\title
{The Evolution of Early-type Field Galaxies Selected from a NICMOS Map
of the Hubble Deep Field North\altaffilmark{1}}

\author{S. A. Stanford\altaffilmark{2}}

\affil{Physics Department, University of California at Davis, Davis,
CA 95616}

\author{Mark Dickinson, Marc Postman, Henry C.\ Ferguson, and Ray A.\ Lucas}

\affil{Space Telescope Science Institute, 3700 San Martin, Baltimore,
MD 21218}

\author{Christopher J.\ Conselice}

\affil{Department of Astronomy, Robinson Hall, California Institute of
Technology, Pasadena, CA 91109}

\author{Tam\'{a}s Budav\'{a}ri}

\affil{Department of Physics and Astronomy, Johns Hopkins University, Baltimore,
MD 21218}

\and

\author{Rachel Somerville}

\affil{Department of Astronomy, University of Michigan, Ann Arbor, MI, 48109}

\altaffiltext{1}{Based on observations with the NASA/ESA
Hubble Space Telescope, obtained at the Space Telescope Science Institute, 
which is operated by the Association of Universities for Research in 
Astronomy, Inc., under NASA contract NAS5-26555.}

\altaffiltext{2}{Institute of Geophysics and Planetary Physics,
Lawrence Livermore National Laboratories, Livermore, CA, 94550;}
\begin{abstract}

The redshift distribution of well-defined samples of distant
early-type galaxies offers a means to test the predictions of
monolithic and hierarchical galaxy formation scenarios.  NICMOS maps
of the entire Hubble Deep Field North in the F110W and F160W filters,
when combined with the available WFPC2 data, allow us to calculate
photometric redshifts and determine the morphological appearance of
galaxies at rest-frame optical wavelengths out to $z \sim 2.5$.  Here
we report results for two subsamples of early-type galaxies, defined
primarily by their morphologies in the F160W band, which were selected
from the NICMOS data down to $\H160_{AB} < 24.0$.  A primary subsample
is defined as the 34 galaxies with early-type galaxy morphologies and
early-type galaxy spectral energy distributions.  The secondary
subsample is defined as those 42 objects which have early-type galaxy
morphologies with non-early type galaxy spectral energy distributions.
The observed redshift distributions of our two early-type samples do
not match that predicted by a monolithic collapse model, which shows
an overabundance at $z > 1.5$.  A $\langle V/V_{max} \rangle$ test
confirms this result.  When the effects of passive luminosity
evolution are included in the calculation, the mean value of $V_{max}$
for the primary sample is $0.22 \pm 0.05$, and $0.31 \pm 0.04$ for all
the early-types.  A hierarchical formation model better matches the
redshift distribution of the HDF-N early-types at $z > 1.5$, but still
does not adequately describe the observed early-types.  The
hierarchical model predicts significantly bluer colors on average than
the observed early-type colors, and underpredicts the observed number
of early-types at $z \sim 2$.

Though the observed redshift distribution of the early-type galaxies
in our HDF-NICMOS sample is better matched by a hierarchical galaxy
formation model, the reliability of this conclusion is tempered by the
restricted sampling area and relatively small number of early-type galaxies
selected by our methods.  For example, our results may be biased by the
way the HDF-N appears to intersect a large scale structure at $z
\sim 1$.  The results of our study underscore the need
for high resolution imaging surveys that cover greater area to similar
depth with similar quality photometry and wavelength coverage.

Though similar in appearance in the $\H160$ data, the primary and
secondary samples are otherwise rather different.  The primary sample
is redder, more luminous, larger, and apparently more massive than the secondary
sample.  Furthermore the secondary sample shows morphologies in the
optical WFPC2 images that are more often similar to late-type galaxies
than is the case for the primary sample.  The bluer secondary sample of early-types 
have a star formation history which can be approximated by a Bruzual \&
Charlot $\tau$ model, or by a galaxy formed at high redshift with a 
small, recent starburst.  Given the differences in their apparent
stellar masses and current luminosities, it would seem unlikely that
the secondary sample could evolve into galaxies of the primary
sample.  

\end{abstract}


\section{Introduction}

An understanding of the evolution of elliptical galaxies continues to
elude both observational astronomers and theoretical astrophysicists.
The debate between two competing theories for the formation and
evolution of elliptical galaxies has been guiding most recent
investigations in this area.  The traditional monolithic collapse
model proposed by, e.g., \citet{els}, \citet{searle73}, and \citet{tg76}
postulates a single burst of star formation at high redshift, followed
by passive stellar evolution.  The newer alternative is based in a
cold dark matter cosmogony, wherein galaxies are assembled
hierarchically over relatively long periods of cosmic time.  A
detailed review of the observational evidence for and against both the
monolithic and the hierarchical scenarios has been presented by \citet{schade99}.

Significant observational effort has been spent in investigating these
two galaxy formation scenarios.  Attempts to uniformly select and
study samples of high redshift early-types in the field have used
selection criteria based on morphology, color, or both.  There is
little evolution in the luminosity function of red galaxies at $z <
0.7$ in the CFRS \citep{lilly95}, which appears to contradict basic
expectations of passive luminosity evolution (PLE) models, wherein
galaxies should be more luminous at higher redshift.  Reasonable
interpretations are that ellipticals assemble late by merging
processes, or that some fraction of distant ellipticals are blue
enough to drop out of color-selected samples.  Indeed, morphologically
defined samples have identified blue field ellipticals 
\citep{schade99,men99,treu02,im02}.  Another possible interpretation of
the CFRS results is that galaxies grow in mass through merging while
simultaneously fading.  The issue of the mass of these blue
early-types was addressed by \citet{im01} who suggest that most
of the blue spheroidals being found in the field at moderate redshifts
are low-mass systems undergoing starbursts, rather than massive
ellipticals.

Several studies using morphological identification of early-types in
optical HST images have found little if any change in the space
density of ellipticals up to $z \sim 1$ \citep{driver98, brinch98,
im99, schade99}.  At $z > 1$, the strong k-correction means that
near-IR data are better suited to making an unbiased census of
early-types.  In the two Hubble Deep Fields (HDF), studies using
ground-based near-IR imaging have found a deficit of red ellipticals
at $z > 1$ \citep{zepf97,franc98,barg99}.  More extensive optical-IR
surveys incorporating morphologies based on data from WFPC2
\citep{im99,men99,abraham99} or NICMOS \citep{treu99} have also
concluded that there are fewer luminous, red ellipticals at $z > 1$
than would be expected from PLE models.  IR surveys of larger areas
which rely on colors instead of morphologies to select early-types
have found a variety of values for the space density of red galaxies
at higher redshifts \citep{mccrack00,daddi00,pat01}.
 
A continuing source of uncertainty in the interpretation of such
surveys is the way that large scale structure may influence the
results given the relatively small areas typically covered in near-IR
imaging surveys.  Our study is particularly affected by this problem
because red galaxies cluster even more strongly than do field galaxies in
general.  \citet{daddi01} found a comoving correlation length
$r_0 h= 12\pm3$ Mpc for a sample of 400 extremely red galaxies ($R - K_s >5$ to $K_s =
19.2$) in a 700 arcmin$^2$ survey.  Another problem when comparing deep fields
such as the HDF to a local galaxy population in order to measure evolution is
the uncertainty in the faint end of the local galaxy luminosity function. 
Since the HDF data reach such faint magnitudes, the majority of
the galaxies have low luminosities, and hence the comparison with the local
population is sensitive to the faint end slope.  Improvements in our knowledge
of the local LF due to the Two Degree Field Survey and the Sloan Digital Sky
Survey should help to alleviate this problem in the near future.  

Due to the extreme depth of its high resolution imaging, the HDF comes into 
its own in the important redshift regime of $1 < z < 2$.  The large observed
wavelength range and accuracy of the combined WFPC2 and NICMOS data makes it
possible to obtain relatively accurate photometric redshifts for those
objects without spectroscopic redshifts.  By making use of our NICMOS map of
the entire HDF in the \J110 and \H160 bands, we can exploit 
a dataset which is unique in being able to study the rest frame
optical properties of high redshift early-type galaxies.   The
available dataset allows us to largely solve problems having to do with the
adequate and consistent selection of early-type galaxies up to redshifts high
enough ($z \sim 3$) to reach their likely formation epoch(s).  Recently,
\citet{dick03} have exploited the advantages of the same dataset to
examine the evolution of the global stellar mass density at $z < 3$.

Though dependent on the details and on the particular cosmology, in general
terms the two galaxy formation scenarios predict significantly different
histories for the evolution of the space density of early-type galaxies. 
Assuming a single formation epoch, the number of E-S0s in a given mass range
remains constant with time in the monolithic model.  In a hierarchical model,
the situation is more complex. Early type galaxies can grow in mass by merging,
or acquire a new disk and become 'late type'. As a general trend, the number of
massive early type galaxies should increase with time in this
scenario.  The space density test is potentially the most powerful in
distinguishing between the two formation scenarios.  There are at least two
significant problems relevant to our investigation in carrying out such an
experiment.  First, we typically measure brightness and redshift to calculate
the luminosity of a galaxy, but do not know the mass.  Currently the best
practical method of overcoming this problem for a large, faint sample is to
determine the luminosity of a galaxy as close in wavelength as possible to the
rest frame near-IR, where the mass-luminosity relation is most stable against
perturbations from dust and recent star formation.  In the future it may become
feasible to obtain a more accurate dynamical measure of the mass for large
numbers of distant galaxies.  Second, we must select early-type galaxies from
observational surveys.  This second problem can be addressed by obtaining the
highest resolution images possible and performing a morphological selection.

Though such a morphological method of selecting early-types may be the
best available, it is not without potential pitfalls.  Even with the
excellent HST angular resolution and deep images of the HDF,
morphological classification of very faint galaxies is difficult.
Selection biases as a function of magnitude, redshift, and galaxy type
are probable.  We describe below tests using simulations in an attempt
to both quantify these biases and to qualitatively understand them.  

Another potential problem with the morphological selection technique
could occur if an early-type galaxy were undergoing an interaction
or were forming during a major merger event.  Then it might not be
classified as an ``early type'' and thus could be excluded from our
sample.  Similarly, the recent advances made by SCUBA in finding
evidence of dust-enshrouded star formation at high redshift also has
complicated understanding of the origin of early-type galaxies.  It
may be that the high-$z$ SCUBA sources will become early-type galaxies,
but that these sources are too heavily obscured to show up in deep
near-IR imaging surveys.  If the antecendents of today's early--type
galaxies cannot be recognized by their morphologies at high redshift,
then it will be difficult to wisely choose adequate samples for
comparison with present--epoch ellipticals.  

This paper presents the results of our investigation of early-type
galaxy evolution, based on deep NICMOS and WFPC2 images of the HDF--N.
We will describe the way early-type
samples may be selected by using morphology and/or on the basis of the colors. 
These samples are then compared with the predictions of monolithic collapse
and of hierarchical models.  Though the results of these comparisons are clear,
their significance to understanding elliptical and S0 galaxy formation and
evolution is tempered by the fact that the HDF-N represents only one very small
window into the Universe, which may be strongly affected by large
scale structure.  Perhaps more valuable are the results
obtained from comparing the subsamples of red and blue early-types
concerning the nature of early-type galaxy evolution.  The
assumed cosmological parameters are $h = H_0 / (100$ km s$^{-1}$ Mpc$^{-1}) =
0.7$, $\Omega_M = 0.3$, and $\Omega_\Lambda = 0.7$.  Over the redshift
range of interest, the linear resolution of the data does not change
by much.  Approximately 90\% of the early-types found in the NICMOS
data span a redshift range such the linear resolution varies by a
factor of only 1.43.  

\section{Data}

The HDF--N was observed by NICMOS between UT 1998
June 13 and June 23, when the HST secondary mirror was at the optimal
focus for Camera 3.  Our observations and data reduction will
be described in detail elsewhere (Dickinson et al.\ in preparation);
the relevant aspects are summarized here.  The complete HDF--N was
mosaiced by Camera 3 with 8 sub--fields, each imaged during three
separate visits.  During each visit, exposures were taken through both
the F110W (1.1$\mu$m) and F160W (1.6$\mu$m) filters.  (Henceforth we
will refer to the six WFPC2 and NICMOS HDF bandpasses used on the
HDF-N as \U300, \B450, \V606, \I814, \J110 and \H160.)  Each section
of the mosaic was dithered through 9 independent positions, with a net
exposure time of 12600s per filter, except in a few cases where
telescope tracking was lost due to HST Fine Guidance Sensor failures.

The data were processed using STScI pipeline routines and custom
software, and were combined into a single mosaic, accurately
registered to the HDF--N WFPC2 images, using the ``drizzling'' method
of \citet{driz}.  The Camera 3 images have a point spread function
(PSF) with FWHM~$\approx$~0\secspt22, primarily limited by the pixel
scale (0\secspt2).  Sensitivity varies over the field of view due to
variations in NICMOS quantum efficiency and exposure time.  On average
the images have a signal--to--noise ratio $S/N \approx 10$ within a
$0\secspt7$ diameter aperture at $AB \approx 26.1$ for both the \J110
and \H160 filters.\footnote{Unless otherwise stated, we use AB
magnitudes throughout this paper, defined as $AB = 31.4 -
2.5\log\langle f_\nu\rangle$, where $\langle f_\nu \rangle$ is the
flux density in nJy averaged over the filter bandpass.}  In order to
ensure properly matched photometry between the optical and infrared
images, the WFPC2 data were convolved to match the NICMOS PSF.
Photometric catalogs were constructed using SExtractor \citep{bert96},
by detecting objects in a sum of the \J110 and \H160 images and then
measuring fluxes through matched apertures in all bands.  Objects were
classified as star, and removed from the catalog, as a result of
spectroscopic information, of having colors similar to stars, of their
point-like appearance in the HST images, and of their having high {\it
CLASS\_STAR} values according to SExtractor.  The complete galaxy
catalog down to \H160 $< 26.0$ will be referred to as the HDF NICMOS
Mosaic, or HNM, catalog.  Information for the $\H160 < 24.0$ sample is
presented in Table 1, including catalog number, the "total" i.e.\ {\it
MAG\_AUTO} magnitude $H_{160}^{k}$, and the isophotal $\I814 - \H160$
color, as well as other parameters to be defined in later sections.
{\it MAG\_AUTO} is the magnitude measured within an ellipse whose size
is defined by radial moments of the light profile \citep{kron80}, and is
designed to enclose a large and roughly constant fraction of
the galaxy light for a variety of surface brightness profiles.

Additional $K$-band photometry was obtained from the imaging carried
out by \citet{dick98} on the HDF-N using IRIM on the KPNO 4~m
telescope.  A version of the fitting procedure described by
\citet{fern99} was implemented by \citet{pap01} to
optimally measure photometry from the ground-based images which had
been matched to the HST images.
 
\section{Sample Definition}

To produce samples of early-type galaxies, we have followed three
methods to select objects from the HNM catalog at \H160 $< 24.0$:
classical Hubble typing by visual inspection, de Vaucouleur function
fitting of surface brightness profiles, and the construction of galaxy
spectral types that result from photometric redshift estimates. The
three early-type selection methods express traditional ideas about the
definition of early-type galaxies, and are described below in more
detail.  

\subsection{Visual Classification}

\begin{figure}
\plotone{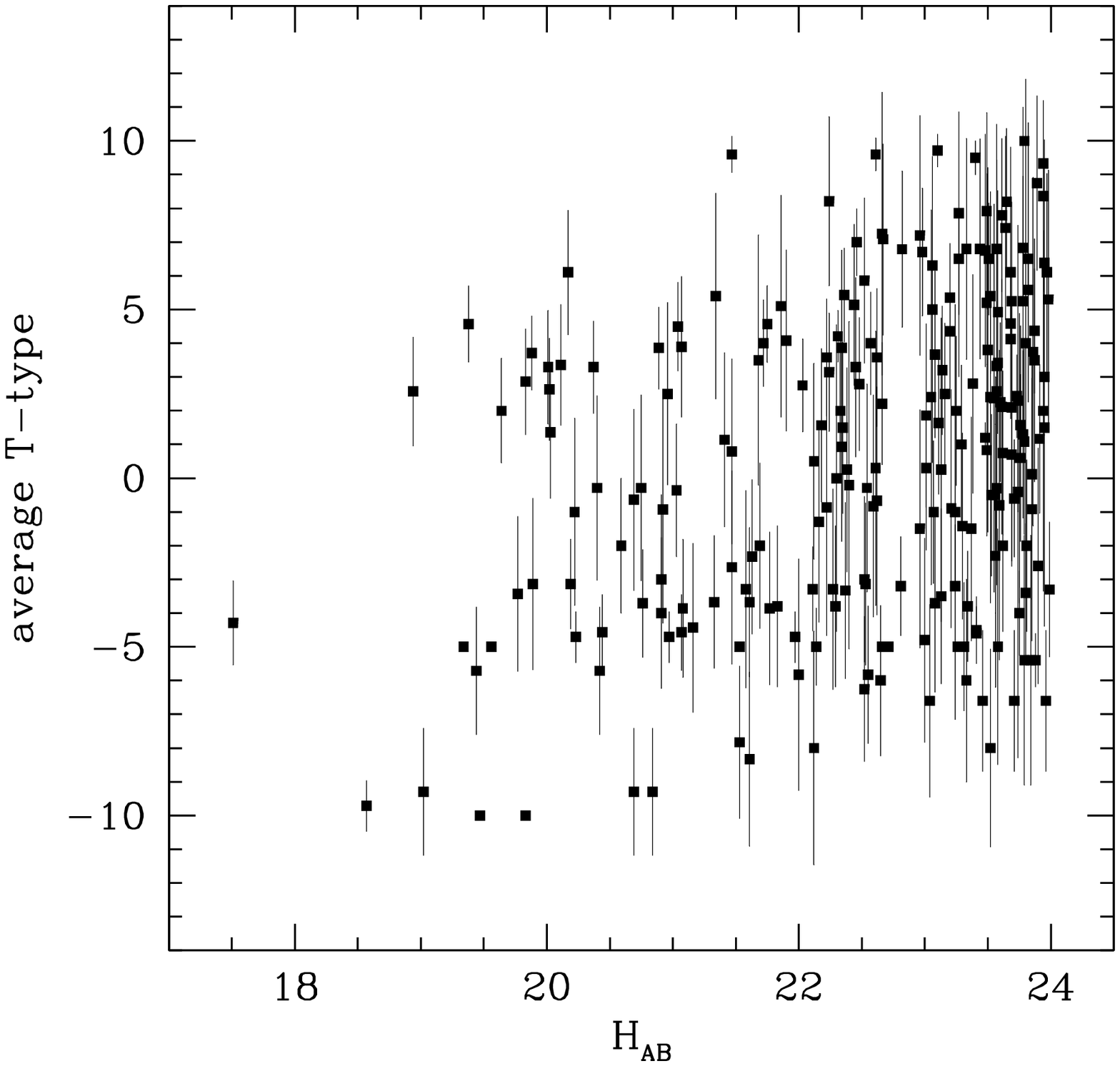}
\caption{Average T-types vs \H160 for $H < 24.0$ sample.  Vertical bars represent the 1
$\sigma$ values on each average. } 
\label{h24_ttvh}
\end{figure}

As a first attempt to obtain basic morphological information on the
objects in the HNM catalog, the \H160 images of all 230 objects with
$\H160 < 24.0$ were visually inspected by 7 members of the NICMOS GO
program team.  Hubble types were assigned by each classifier for each
object where possible.  The typing became difficult for the smallest
objects; some classifiers did not assign types in these cases.  The
Hubble types were converted to T-types (hereinafter TT).  The T-types for
common Hubble types are as follows: $E = -5, S0 = -2, Sa = 1, Sb = 3, Sc = 6$,
and $Im = 10$ (RC3).  Objects which were found to be too compact for
classification were given a TT of $-$10.  The T-types were averaged and the rms
calculated.  Objects for which the rms of the TT was relatively large (greater
than $\sim$3) were reconsidered.  In most of these cases, the T-type assigned
by one classifier was significantly different from the other values.  In these
cases, the discrepant value was discarded, the T-type recalculated and compared
with the visual appearance in all of the available HST bands.  In some of these
cases, the assigned T-types were distributed over a large range, accounting for
the large rms on the average.  These T-types could not be reconciled and so the
original average value is retained along with its large rms, signifying a very
uncertain classification.  The average T-types are plotted against their \H160
magnitudes in Figure~\ref{h24_ttvh}, and are listed in Table 1 along with their
uncertainties.  Images of all of the objects in this catalog may be viewed by
accessing the website www.stsci.edu/$\sim$med/hdfnic3.

\begin{figure}
\plotone{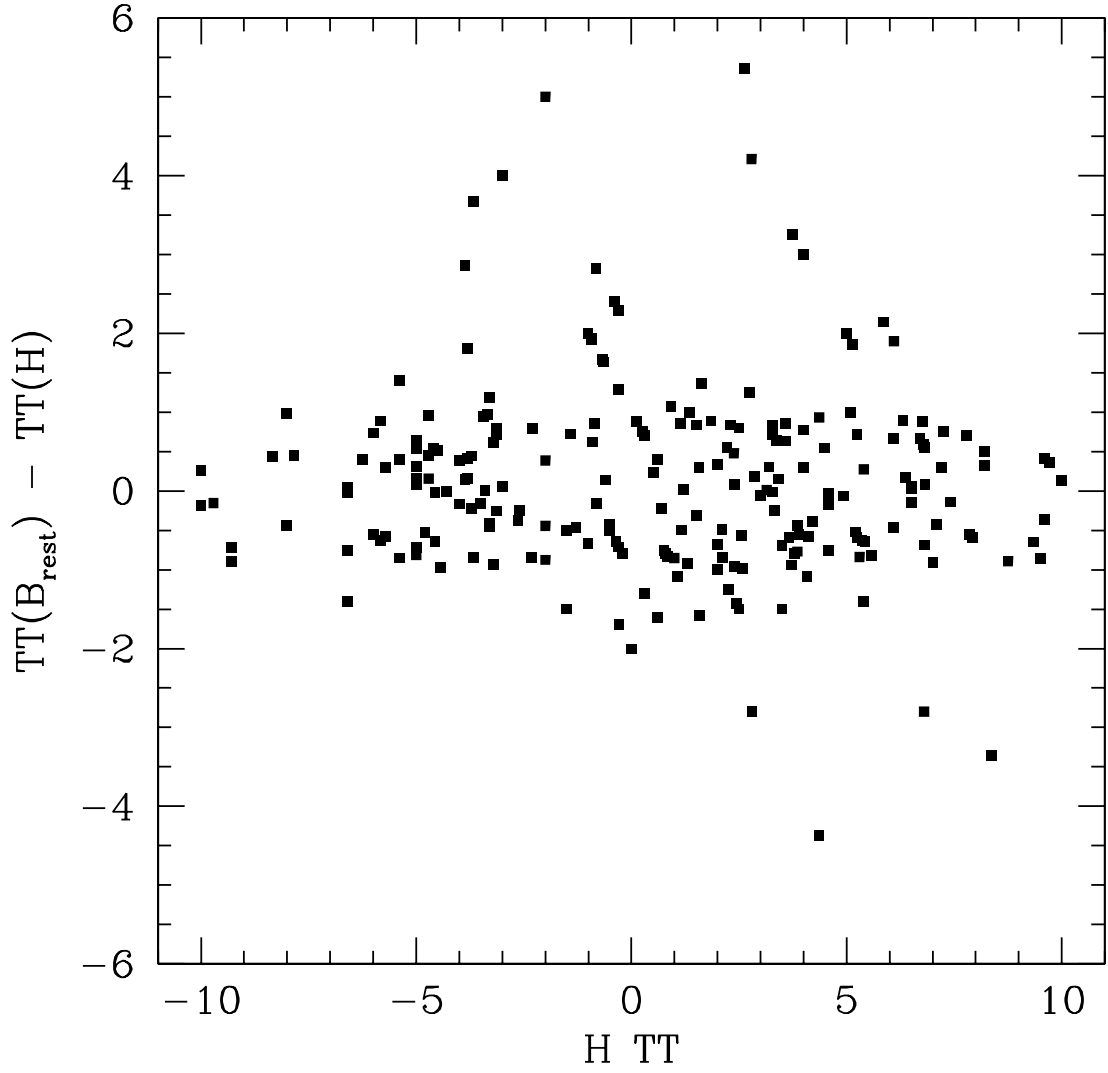}
\caption{The difference TT($B_r$) - TT(\H160) vs TT(H) for $H < 24.0$
sample. }
\label{h24_ttcomp}
\end{figure}

\begin{figure}
\plotone{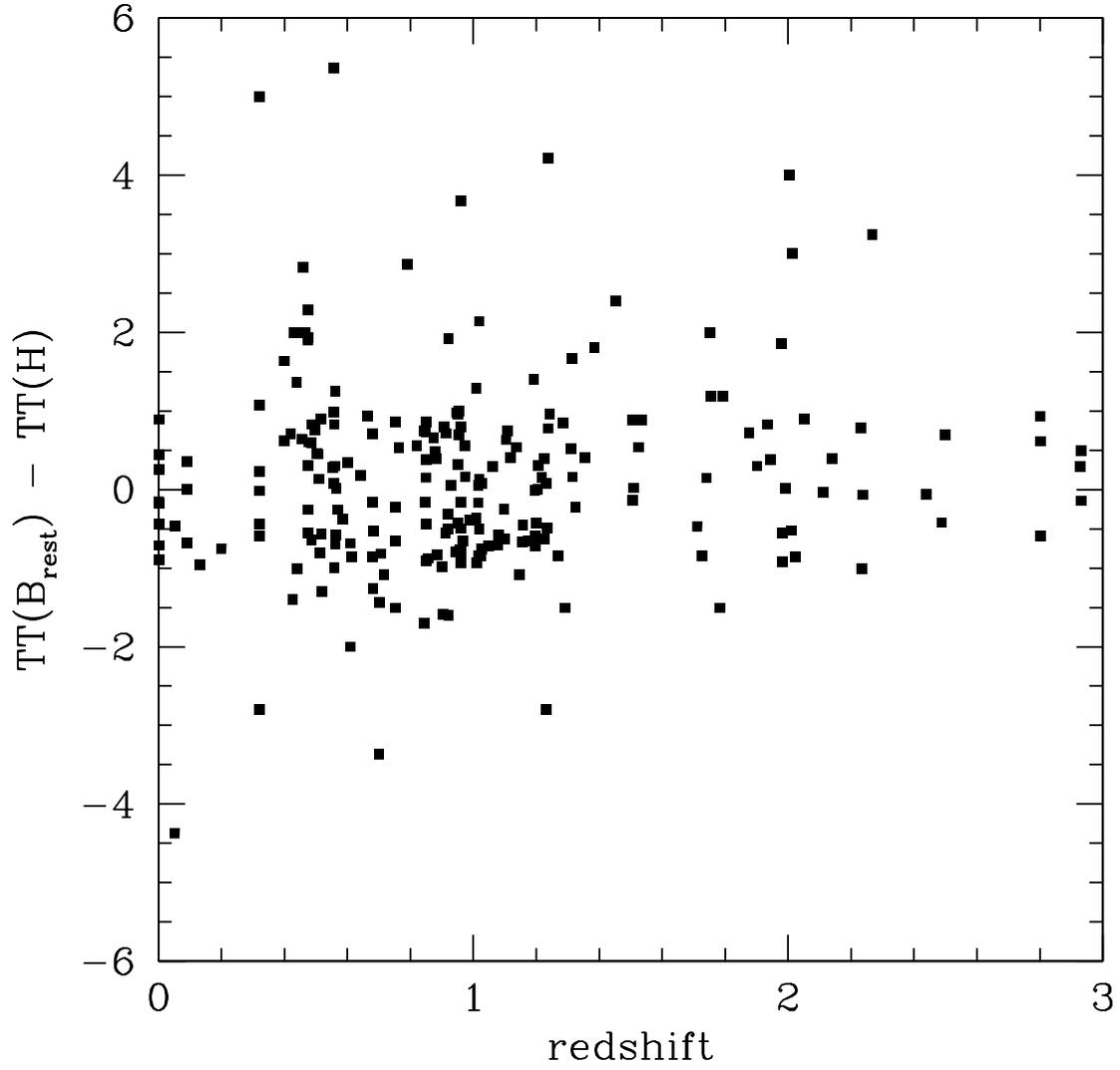}
\caption{The difference TT($B_r$) - TT(\H160) vs redshift for $H < 24.0$
sample. }
\label{h24_ttzcomp}
\end{figure}

Another visual classification was also performed by one of the authors
(SAS).  The $HST$ band which most closely samples the rest frame
$B$-band, called $B_r$, for each object was used to determine its
T-type.  The redshifts used for this procedure were spectroscopic
where available; otherwise photometric redshifts (see Section 3.3)
were used.  The highest redshift that could be typed in this manner
was $z \sim 3$.  Beyond this redshift, the necessary observed band has
a wavelength redder than the available $HST$ imaging.  In general, the
$B_r$ T-types agree with the average values determined from the
observed \H160 image.  In Figure~\ref{h24_ttcomp}, the difference in
T-types determined from the \H160 and $B_r$ images is shown as a
function of \H160 T-type.   Little systematic trend is
evident.   Similarly, in Figure~\ref{h24_ttzcomp}, the redshift
distribution of the TT differences does not show a systematic trend
with redshift.   The TT determined from the $\H160$ band is used for
all morphological types in the rest of this work.  

\begin{figure}
\plotone{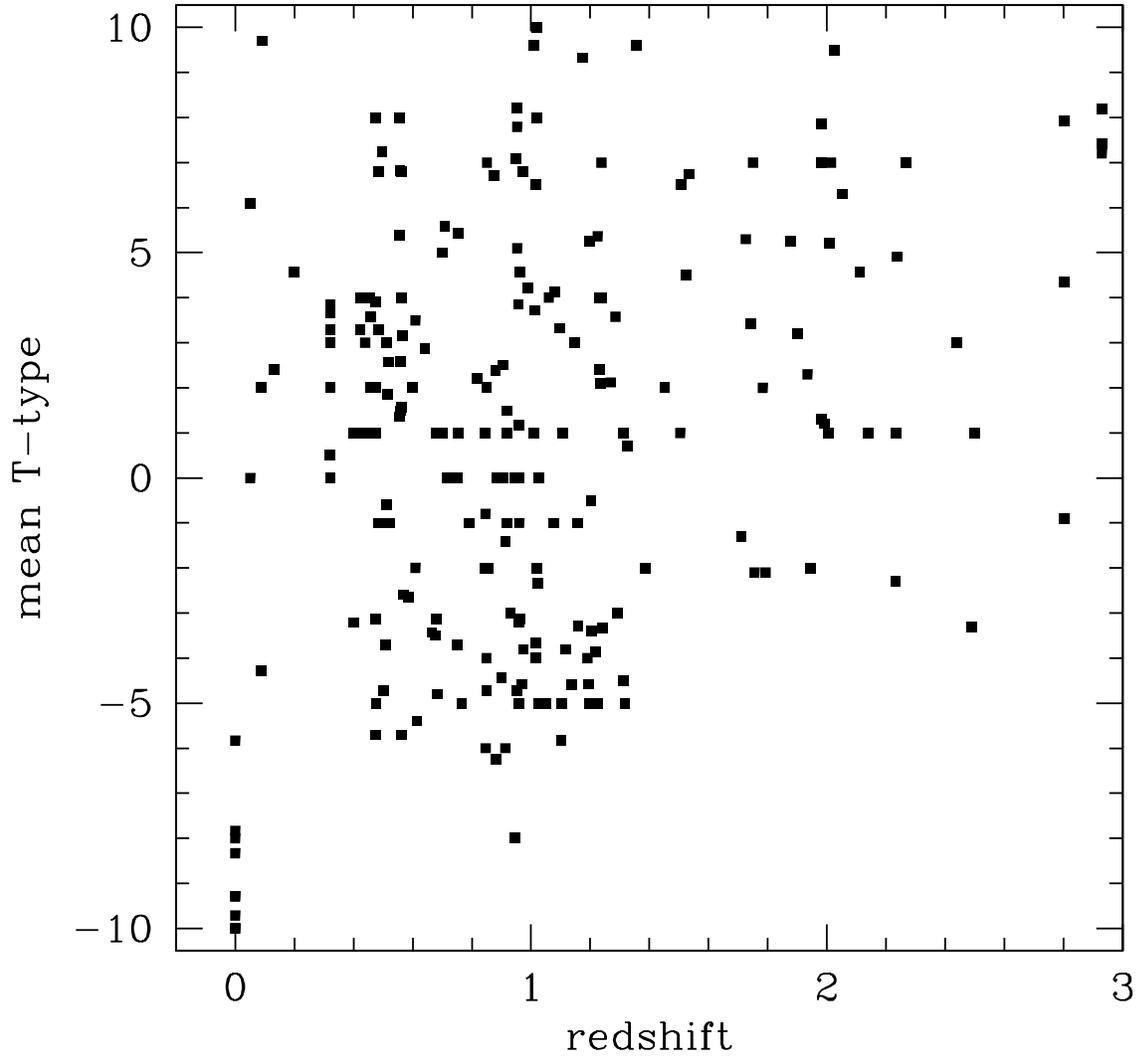}
\caption{Average T-types vs redshift for $H < 24.0$ sample.  The upper
$z$ limit is set by the available TT($B_r$).} 
\label{h24_ttvz}
\end{figure}

To determine the biases in the visual classifications, we performed
the following test.  The $B$ and $g$ images from the \citet{frei96}
sample of nearby galaxies were transformed so as to appear to be
$L^\ast$ galaxies at $z=0.75$ with the S/N of our data and then
convolved with the NIC3 F160W PSF.  These images were then visually
classified by four of the authors and a comparison made with the known
T-types from the RC3.  This test found a tendency to classify galaxies
towards earlier types by $\sim$0.7 T-types on average, with the rms on
the average difference between the classified and real T-type being
$\sim$2.8.  Because of surface brightness dimming, it would be natural
to suspect that we would preferentially assign earlier types to higher
redshift objects if the fainter disks are harder to see with
increasing $z$.  While our test suggests that this has not happened
with our classifications of the NIC3 images, it is not possible to
rule out the surface brightness dimming bias completely.  The
distribution of assigned T-type against redshift is shown in
Figure~\ref{h24_ttvz}, where we see that higher redshift objects have
a slight tendency to have been given later T-types.  However we do not
know the intrinsic distribution of T-types as a function of redshift.
Indeed even though the HST resolution is high, the amount of resolved
detail in physical terms in the highest redshift objects is still less
than for the lower redshift objects.  So it would be likely for the
assigned morphologies of the highest redshift objects to be biased.
Probably even more important in the context of these tests is that the
lower spatial resolution of the NIC3 camera significantly reduces the
apparent contrast of structural features such as spiral arms.

\begin{figure}
\plotone{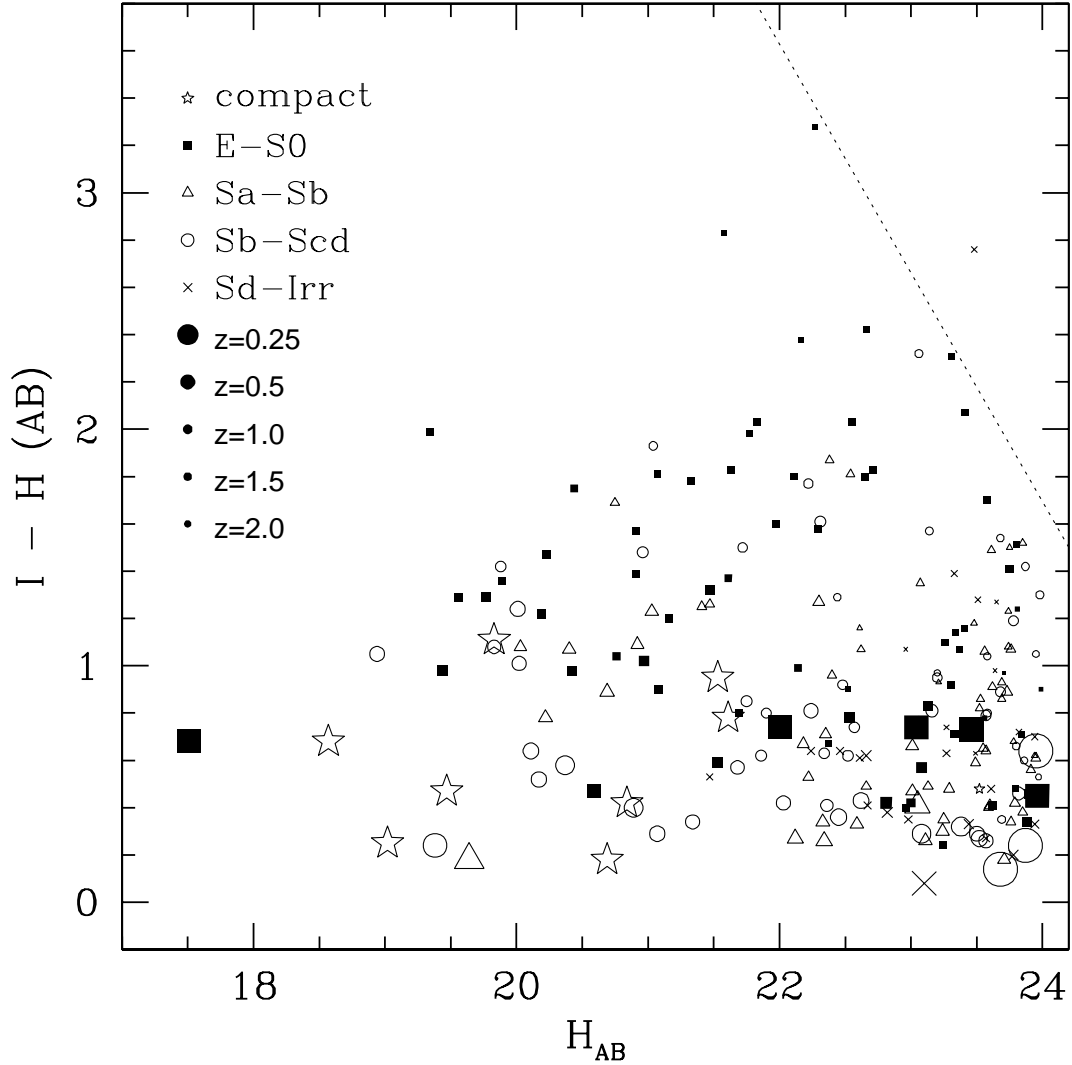}
\caption{$I-H$ vs \H160 for $H < 24.0$ sample.  The point symbol types
denote the Hubble types as indicated in the legend.  Objects with
assigned T-types $< -7$ are termed as ``compact''; these include
stars. The diagonal dotted line indicates the 5 $\sigma$ limit in the
color. The point size is inversely proportional to the redshift of the
object, as indicated in the legend.}
\label{h24_imhvh}
\end{figure}

The distribution of Hubble types, based on the assigned T-types, is
shown for the $\H160 < 24.0$ sample in a color magnitude diagram in
Figure~\ref{h24_imhvh}.  If early-type galaxies are defined as those
objects with $-7 \lesssim $T-type$ \lesssim -2$, then there are 66
such galaxies in the HNM catalog down to \H160 $ = 24.0$.

\subsection{Profile Fitting}

\begin{figure}
\plotone{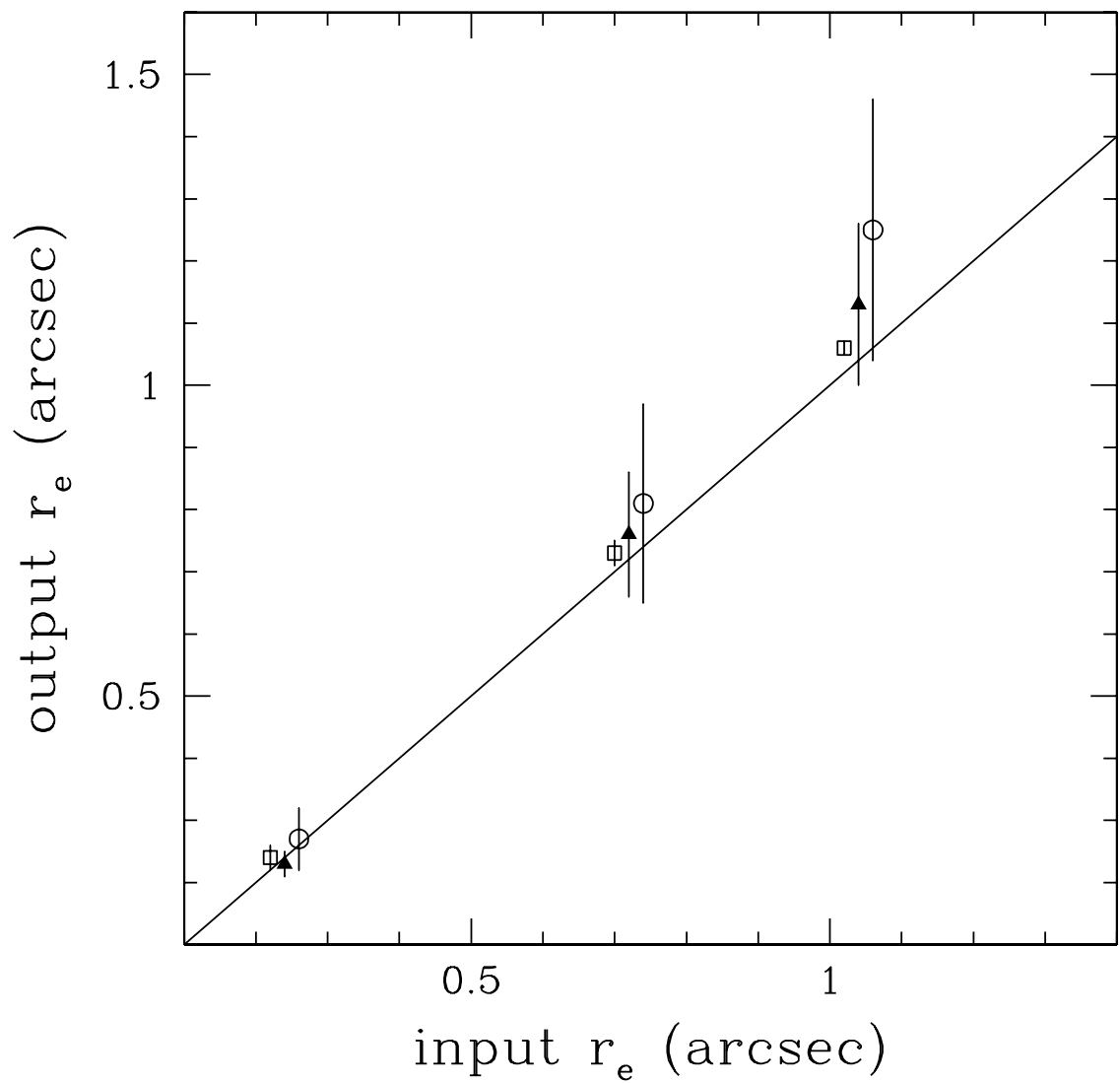}
\caption{A plot of the $r_e$ determined by our profile fitting vs the
input $r_e$ for two sets of model elliptical galaxies.  The open
squares represent models with $\H160 = 20.0$, the solid triangles
those with $\H160 = 22.0$, and the open circles those with $\H160 =
24.0$.  The points have been offset slightly in the horizontal
direction for clarity.  Errorbars representing one $\sigma$ are
shown on each point. }
\label{resim}
\end{figure}

Another method was employed to define early-type
galaxies in the HNM sample by using profile fitting.  Elliptical
isophotes were fit to the \H160 images of each object.  The center of
each ellipse was held fixed while the position angle and ellipticity
were allowed to vary with semimajor axis during the fitting.  The
resulting surface brightness profile was fit by a convolution of a de
Vaucouleur's law with the NIC3 F160W PSF.  The fitting was done over a
range in radius of $0\farcs08 < r < 2\farcs0$ and the model with the
lowest $\chi^2$ was chosen as being the best fit.  Simulations were
performed of this procedure, using model galaxies with de Vaucouleur
profiles and known $r_e$, to determine how well $r_e$ could be
determined.  These models included the effects of the background level and
the PSF.  As shown in Figure~\ref{resim}, over the range $0\farcs3 < r_e <
1\farcs1$ and $24 < \H160 < 20$, $r_e$ is found to vary within 25\% of the
input value using our fitting parameters.

\begin{figure}
\figurenum{7}
\plotone{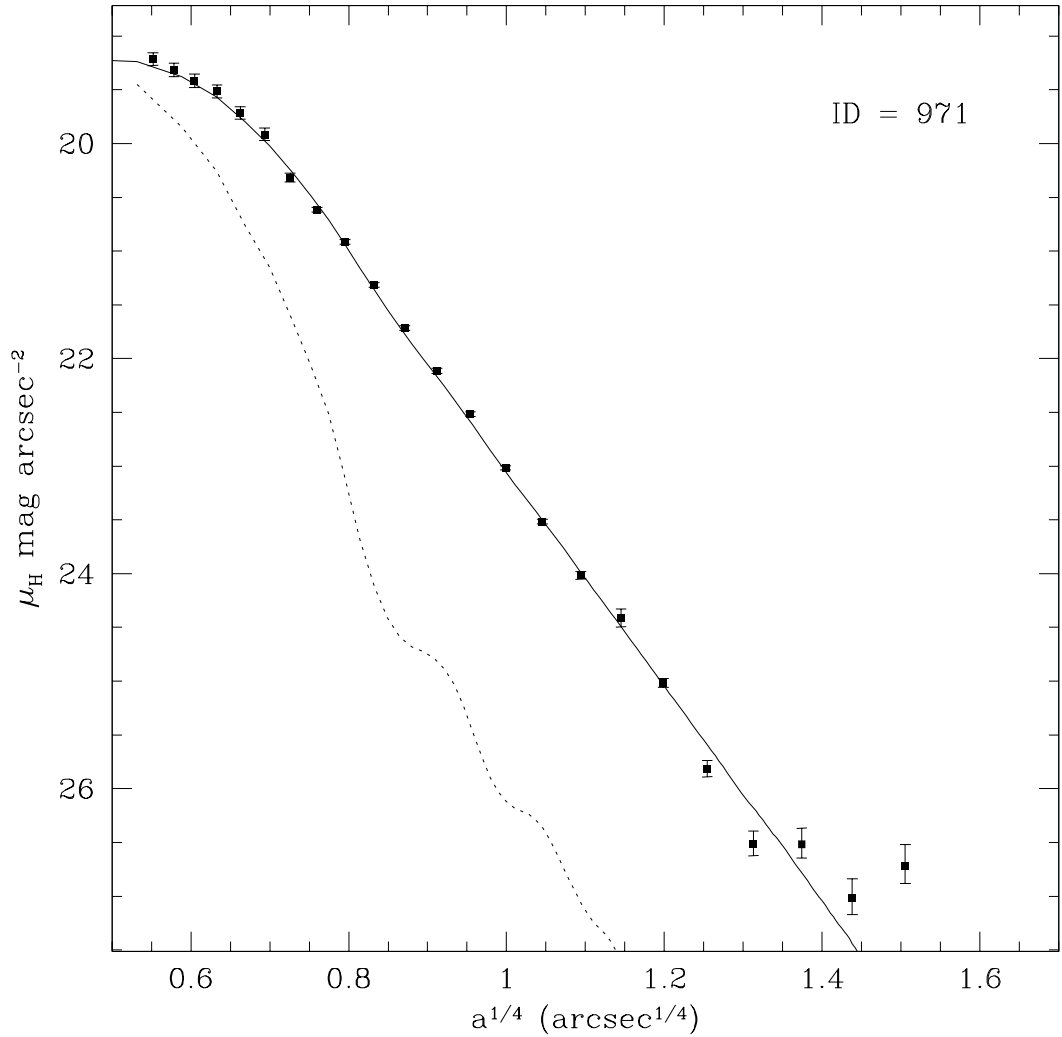}
\caption{Examples of the quality of $r^{1/4}$ law fits to the \H160
surface brightness profiles.  In each plot, the object profile is
represented by the solid square points with one $\sigma$ errorbars,
the F160W PSF by a dotted line, and the best fitting $r^{1/4}$ law,
convolved with the PSF, by a solid line. The object ID number,
corresponding to the entries in Table 1, are shown in each panel.
Here a good fit is shown. }
\label{profiles}
\end{figure}

The best fit $r_e$, $\mu_e$ (the half-light radius and the average surface
brightness within that radius) were recorded along with a visual estimate
of the quality of the fit for each galaxy in the $\H160 < 24.0$
sample.  These quality estimates are given in the column labelled
``q'' in Table 1, with p representing poor, f for fair, and g for
good.  These qualitative classifications were deemed adequate because
the profile fitting is not being used as the main method of selecting
the early-type galaxies.    In cases where the object is obviously e.g.\ an irregular galaxy,
the fits are essentially meaningless; these are classified as poor
fits.  For some objects, the quality of the fit to the surface
brightness profile by the $r^{1/4}$ law is fair but not good enough to
exclude a reasonable fit by an exponential profile.  The objects where
the fit to the surface brightness profile is good may be considered to
be early-type galaxies; there are 52 such objects in the HNM down to
$\H160 < 24.0$.  Examples of the profile fits are shown in
Figure~\ref{profiles}.  

\subsection{Photometric Redshifts and Spectroscopic Types}

\begin{figure}
\figurenum{8}
\plotone{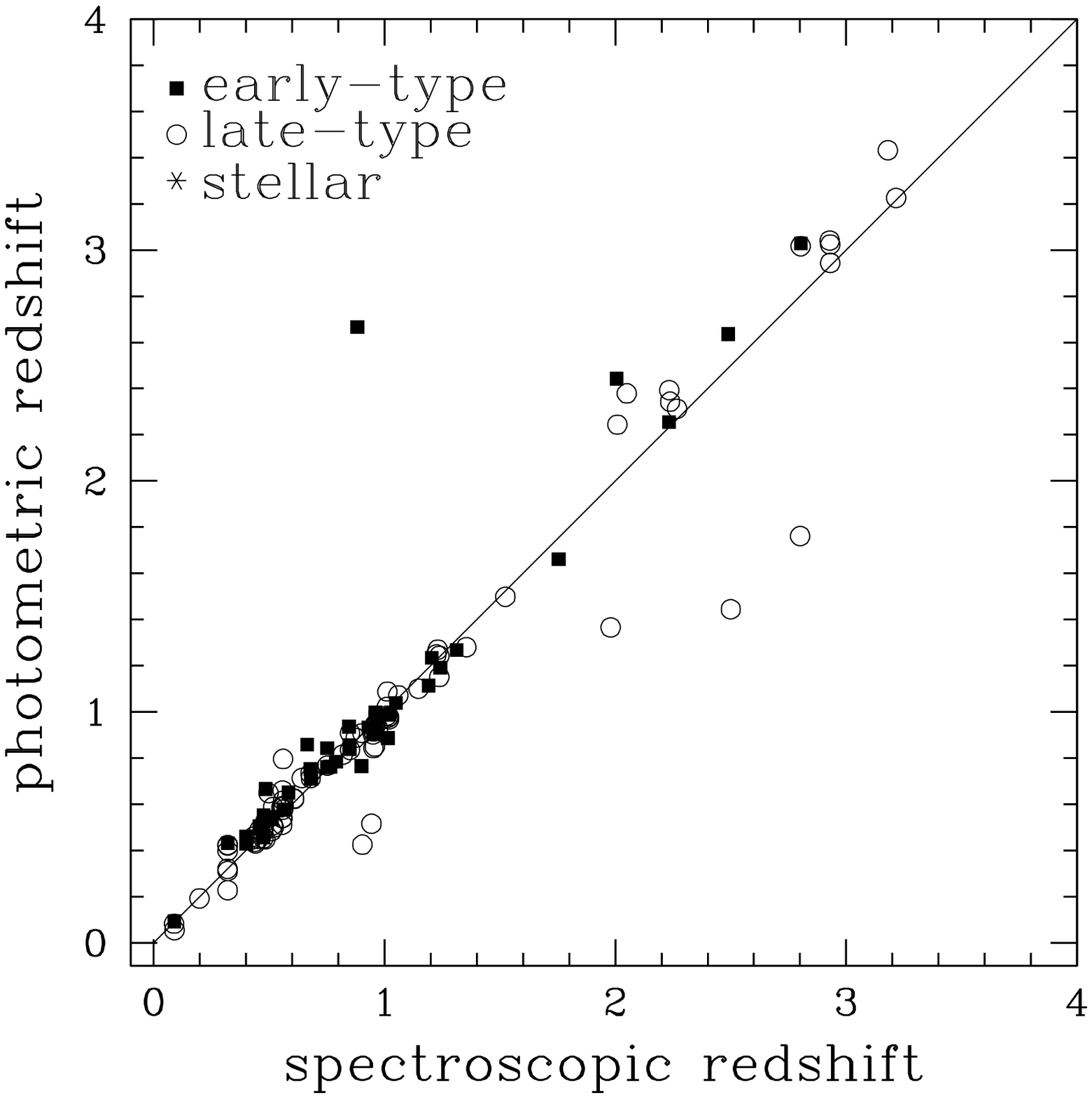}
\caption{Photometric vs spectroscopic redshifts where the latter are
available.  Morphologically-defined early and late-type galaxies are shown by
different symbols.  Errorbars are not shown to improve clarity since
they are almost never larger than the points.}  \label{h24_zpvzs}
\end{figure}

Spectroscopic redshifts are used where available and are listed in
Table 1.  These have been taken from the literature, mostly taken from
\citet{cohen00,cohen01}, and \citet{dawson01} and references therein,
along with a few redshifts made available by C.\ Steidel and K.\
Adelberger (private communication).  For those objects without
spectroscopic redshifts, photometric redshift estimates were obtained
from \citet{bud00} who used all 7 bands for which we have photometry;
the 6 HST bands plus the $K$-band data from KPNO.  A comparison
between the best-fit photometric $z$ and the spectroscopic redshifts,
where available, may be seen in Figure~\ref{h24_zpvzs}.  As described
fully in \citet{bud00}, the best fitting combination of 3 eigenspectra
from \citet{ccw} at the most likely redshift is used to obtain the
galaxy spectral type (hereinafter the S-type or ST), for each object,
which is listed in Table 1, along with the photometric redshift.  For
our purposes we have chosen to define early-type galaxies as having
spectral types less than 0.2. We decided to use a higher value of 0.2,
compared to the value of 0.1 given by \citet{bud00}, to define an
early-type SED because a higher value allows for some evolution in the
SED towards bluer colors at higher redshifts which has been found to
occur in early-type galaxies \citep{sed98}.  An early-type ST is a
galaxy whose SED is dominated by an old stellar population, which is
not necessarily the same thing as a galaxy with an early-type
morphology.  We will make use of these two definitions of what
constitutes an early-type galaxy below in selecting different kinds of
early-type samples from the HNM.  For reference Sa, Sc, and Irr
galaxies have ST values of 0.15, 0.35, and 0.55 respectively.  If
galaxy spectral type is the only criterion, 54 galaxies with S-type
values less than 0.2 would be selected from the HNM sample down to
$\H160 = 24.0$.

\subsection{Primary and Secondary Subsamples}

\begin{figure}
\figurenum{9}
\plotone{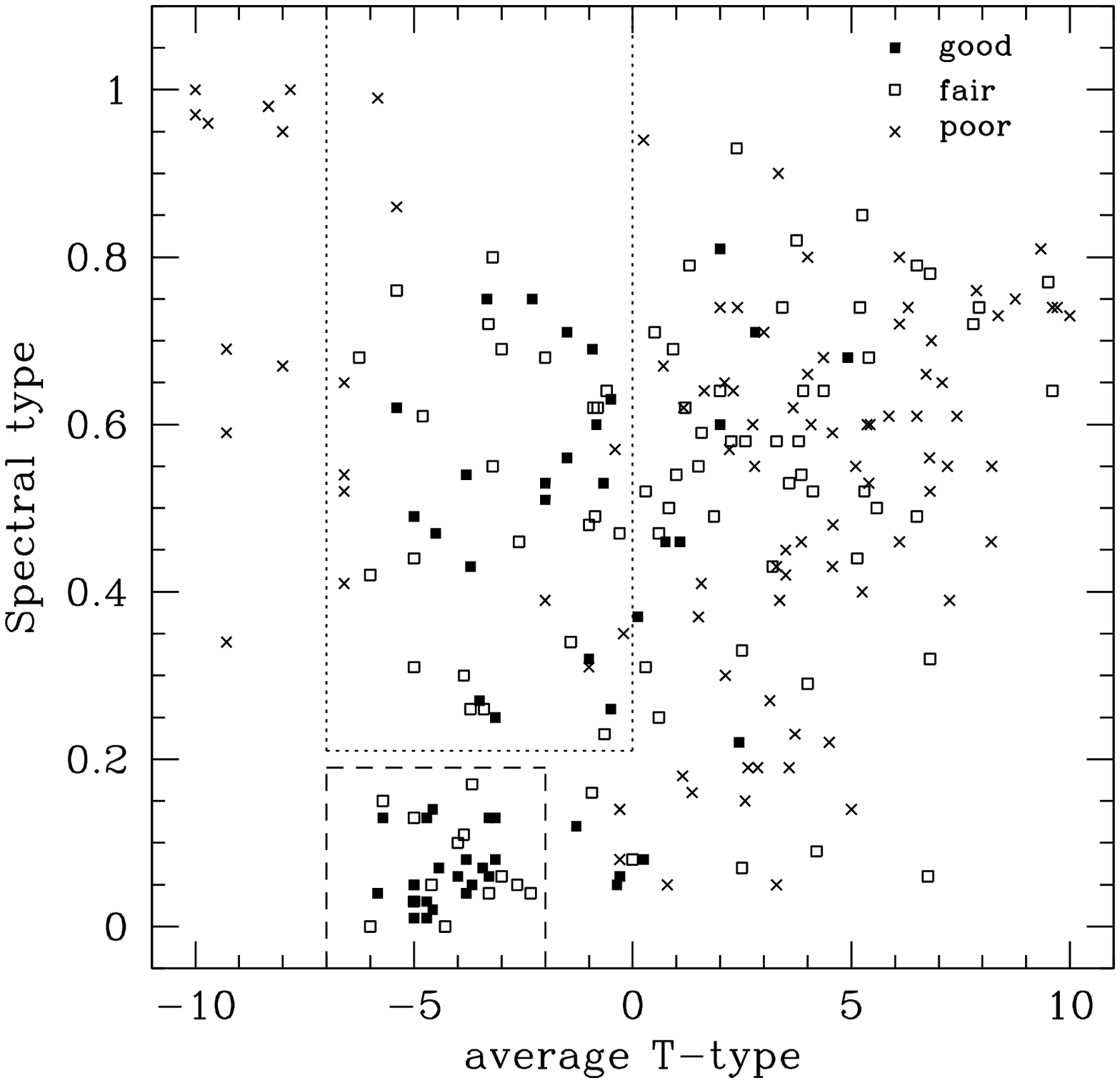}
\caption{Galaxy spectral types, as determined from the photometric
redshift procedure, vs T-types for HNM sample.  The point style
indicates the quality of the de Vaucouleur law fit to the object's
\H160 surface brightness profile, where a solid square is good, an
open square is fair, and a $\times$ is poor.  The dashed area in the
shape of a square near the bottom shows the area of the primary
early-type sample, and the dotted
rectangle encloses the secondary sample of early-type
galaxies, as described in the text.}
\label{h24_stvtt}
\end{figure}

A useful way to compare the three ways of defining early-type galaxies
is shown in Figure~\ref{h24_stvtt}.  The galaxy spectral types are
plotted against the T-types, with the quality of the de Vaucouleur fit
being indicated by the point style.  A very broad correlation is seen
from the combination of earlier T-type and earlier S-type to later
T-type and later S-types.  But there is a large amount of scatter in
the other parameter at a given T-type or S-type.  Nominally the S-type
value should be less than 0.1 to qualify as an elliptical or S0
galaxy, according to the photometric redshift methodology, and the
T-type should be less than $\sim$0.  Figure~\ref{h24_stvtt} shows a
well-defined group of galaxies with $-7 < $ T-type $ < -2$ but with S-types
reaching as high as $ < 0.2$; note that all of these galaxies have
fair to good $r^{1/4}$ law fits.  This combined definition may be a
good way to isolate true early-type galaxies, though it is not without
caveats.  The objects with S-types at $0.1 - 0.2$ may be better fit by
the photo-z technique using later-type spectra because their dominant
stellar populations are younger than a present-epoch elliptical---as
would be expected at higher redshifts.  Furthermore, some of the
objects in this group have profiles which are fit by exponential
profiles as well as by the de Vaucouleur law, indicating a significant
contribution from a disk component to the observed \H160 light is possible.
Finally, there are a significant number of galaxies outside of this
group in Figure~\ref{h24_stvtt} which qualify as early-type galaxies
in at least two of the three methods.  Only selecting objects with
S-types less than 0.1 would add few objects to the group of apparent early-types,
principally with later T-types and a mixture of profile fit qualities.
Selecting galaxies just by T-type would add a significant number of
apparent early-types spanning a large range in spectral type, with
mostly fair to good $r^{1/4}$ law fits.  Selecting galaxies just by
their profile fits would add mainly the same objects as the T-type
selector.  In short, the T-type and profile fitting methods agree the
best amongst the various combinations; this is not surprising, given
that they are morphological selectors, while the S-type has to do more
with stellar evolution.  

We choose to define the 34 galaxies in Figure~\ref{h24_stvtt} with $-7
< $ T-type $ < -2$ and $0 < $ S-type $ < 0.2$, all of which have fair
to good $r^{1/4}$ law fits, as the primary sample of early-types in the
HNM catalog at $\H160 < 24.0$.  16 of these 34 galaxies have
spectroscopic redshifts.  Also, we define a secondary early-type
sample as those 42 objects outside of the primary early-type area in
Figure~\ref{h24_stvtt}, which have $-7 < $ T-type $ < 0$ and fair or
good $r^{1/4}$ law fits.  21 of these 42 galaxies have spectroscopic
redshifts.  Montages of color images of all galaxies in both the
primary and secondary samples are shown in Figure~\ref{primon1} and
Figure~\ref{secmon1}, respectively.  For each galaxy, two images are
shown; one based on a combination of the $IJH$ bands and the other on
the $BVI$ bands.  For both the primary and secondary samples, the
images made from the redder bands generally appear to be early-type
galaxies.  However, the bluer band images of the secondary sample sometimes
show evidence of late-type galaxy morphology.   

\begin{figure}
\figurenum{12}
\plotone{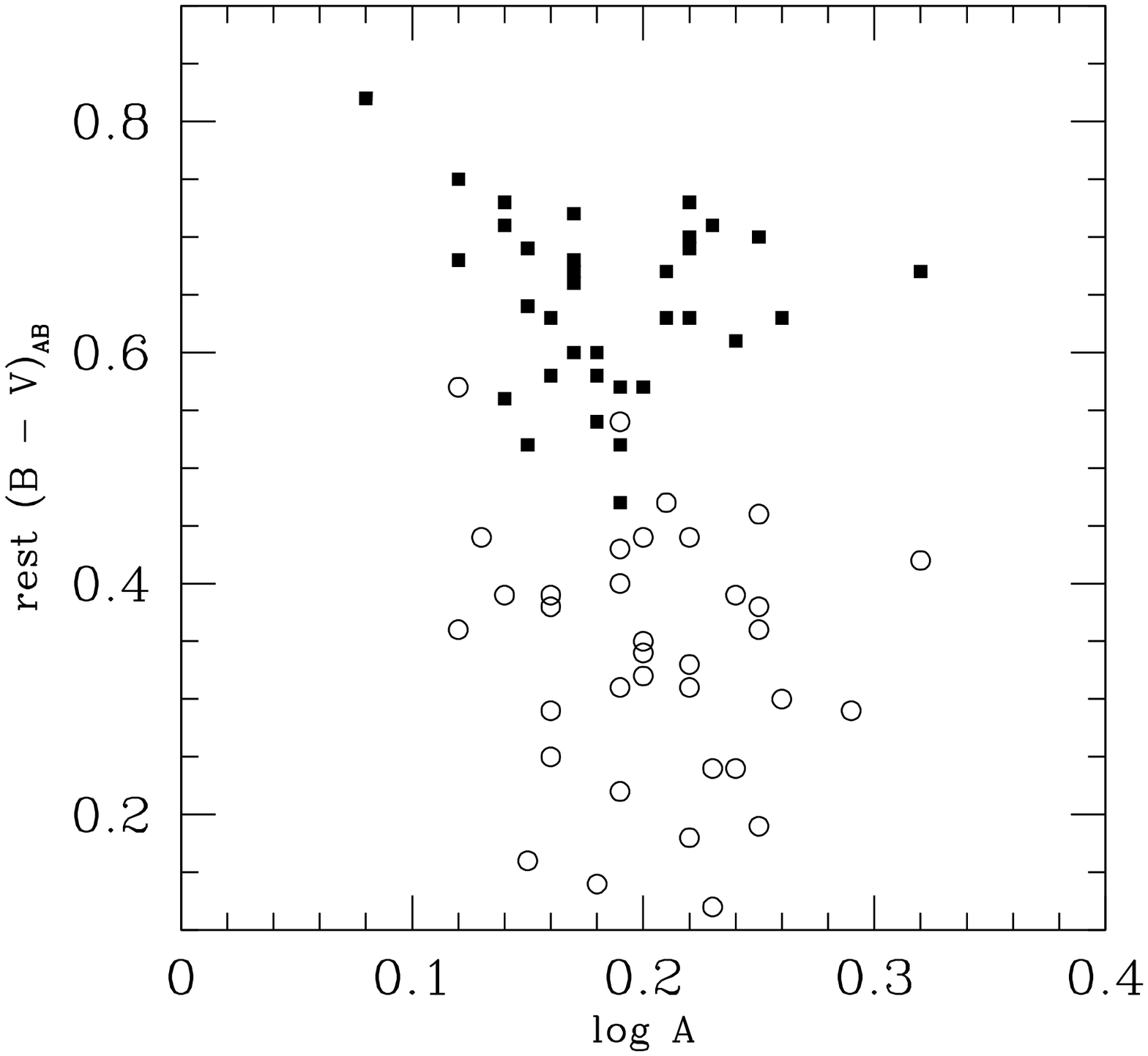}
\caption{The rest frame $B-V$ color for the primary (solid squares)
and secondary (open circles) samples of early-types against the
asymetry index. }
\label{h24_eso_bmvva}
\end{figure}

The F160W images of the galaxies in the primary and secondary samples
were examined to look for the incidence of close neighbors.  One can
imagine that the bluer colors of the secondary early-types could be
due to galaxy interactions.  We found no difference in the frequency
of close neighbors between the two samples.  There is a greater
frequency of poor $r^{1/4}$ fits among the galaxies in the secondary
sample relative to the primary sample.  A plot of the rest frame $B-V$
color against the asymmetry index $A$ \citep{con03} in
Figure~\ref{h24_eso_bmvva} shows only a small difference in the
distributions in $A$ of the two early-type samples, with the secondary
sample having a slightly higher mean value of $A$.

The combined primary and secondary samples form a set of early-type
galaxies selected by morphology.   One of the main goals
of this paper is to understand the relationship between these two
kinds of early-type galaxies.

\section{Detection Simulations}

To determine if and how we are affected by biases
in detecting early type galaxies at high redshifts, we perform two
different types of simulations. In the first, we simulate how 16
nearby giant elliptical galaxies would appear at $z \sim 2$ if
observed under the same conditions in which the NICMOS HDF images were
taken.  We use the B-band images of NGC 2768, NGC 2775, NGC
4125, NGC 4365, NGC 4374, NGC 4406, NGC 4429, NGC 4442, NGC 4472, NGC
4486, NGC 4526, NGC 4621, NGC 4636, NGC 4754, NGC 5322 and NGC 5813
from the Frei et al.\ (1996) sample to perform these simulations.  These
galaxies cover the absolute magnitude range $-23 < M(B) < -19$. K-corrections
were applied to the simulated galaxies, but no correction was made for
evolution.  The luminosity and physical size of each galaxy were preserved 
during the transformation to $z = 2$.  The galaxy images were convolved with
the NIC3 PSF and placed into an image that mimics the actual F160W image of the
HDF-N in terms of sky level and photon noise.  SExtractor was then run on this
image.  We find that we easily detect all 16 of the simulated galaxies.

We perform a similar simulation in which we take all the galaxies in
the Hubble Deep Field with M$_{\rm B} < -18$ and $0.4 < z < 0.8$ and
place them at higher redshifts as they would be observed under the
conditions in which the NICMOS data was taken.  To make the simulated
galaxies, the NIC3 PSF was used.  When these galaxies are placed at $z
\sim 2$ into an artificial NICMOS F160W band image which mimics the
real NIC3 image of the HDF-F in terms of the sky level and noise, and
then re-detected with the same SExtractor detection criteria used on
the original \H160 band images, we find 100\% are retrieved.  In fact,
the detection completeness does not begin to drop until $z \sim 2.5$,
a much higher redshift than that of the decline in the observed
ellipticals.  We therefore conclude that it is unlikely that we are
missing elliptical galaxies at $z > 1.5$ due to our inability to
either detect or identify these systems if they were present.

\section{Evolution of Early-type Galaxies}

We will examine the evolution of the early-type samples described
above in two ways.  First, the spectral evolution of the primary and
secondary samples will be compared and contrasted.  Second, the space
density as a function of redshift will be compared with the predictions of a
monolithic formation model and of an hierarchical model.

\subsection{Spectral Evolution}

\begin{figure}
\figurenum{13}
\plotone{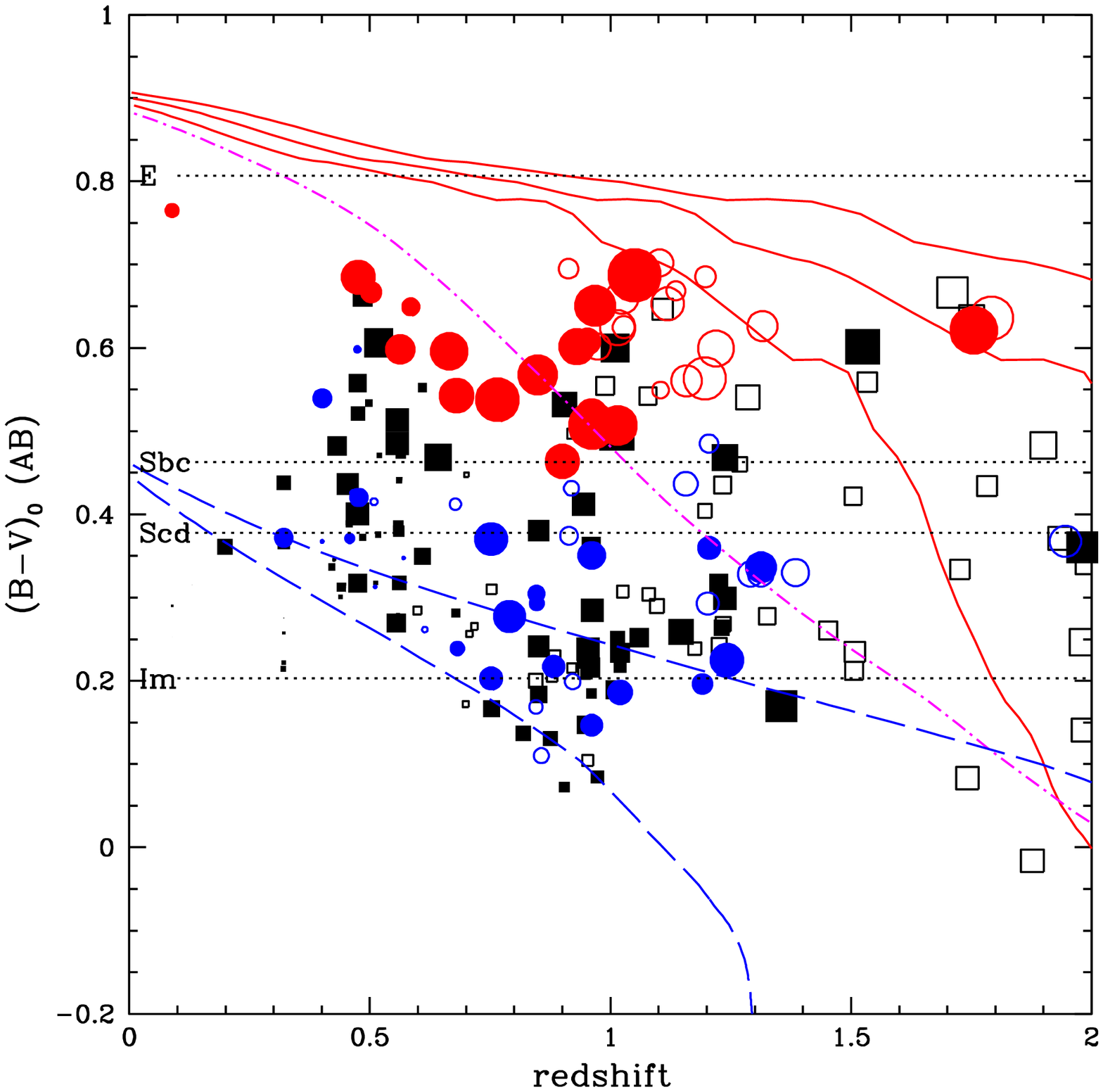}
\caption{Rest frame $B-V$ colors against redshift.  Red circles are
primary early-types, and blue circles are secondary early-types.  The
squares are all other galaxies in the $\H160 < 24.0$ sample.  Filled
symbols have spectroscopic redshifts and open symbols have photometric
redshifts only.  The horizontal dashed lines are the colors of CWW
spectral templates. The curves represent the colors of Bruzual \&
Charlot models described in the text.  The solid red curves represent
a 0.1 Gyr burst which starts at $z_f = 5, 3, 2.1$ followed by passive
luminosity evolution.  The magenta dot-dash line is a $\tau = 1$ Gyr
exponential model with $z_f = 2.5$, and the two blue, long-dash lines
are $\tau = 5$ Gyr exponential models with $z_f = 1.5$ and $z_f = 5$. }
\label{BmV0_vs_z_egals}
\end{figure}

One of the traditional means of analyzing the evolution of galaxy
populations is through the comparison of their colors with spectral
synthesis models.  We make one comparison in rest frame $B-V$ so as to
consistently sample the same part of a galaxy's rest frame spectrum.
The rest frame colors for the objects were obtained by interpolating
among the appropriate observed frame magnitudes as indicated by the
redshift.  Figure~\ref{BmV0_vs_z_egals} shows the rest frame $B-V$
colors of all the $H<24$ sample with $M(V) < -17$, where the symbol
size scales with $M(V)$.  We have calculated the color evolution for a
number of Bruzual \& Charlot (1998 version) models with varying star
formation histories, assuming only $Z_\odot$ stars and shown these in
Figure~\ref{BmV0_vs_z_egals}.  Most of the early-type galaxies lie
between the single-burst $z_f = 5$ and the $\tau$ models, indicating a
wide range in star formation histories.  Lower
metallicity single burst models would yield bluer colors which could
match the colors of the lower-redshift early-types that follow the
$\tau$ model.  Also, it would take only very small starbursts, in
terms of the stellar mass, to make the high redshift $z_f$ PLE models
blue enough to match the colors, if only for a short time, of the
secondary early-types.  Thus, the possibility that the early-types in
the secondary sample formed the vast majority of their stars at high
redshift cannot be excluded.

\begin{figure}
\figurenum{14}
\plotone{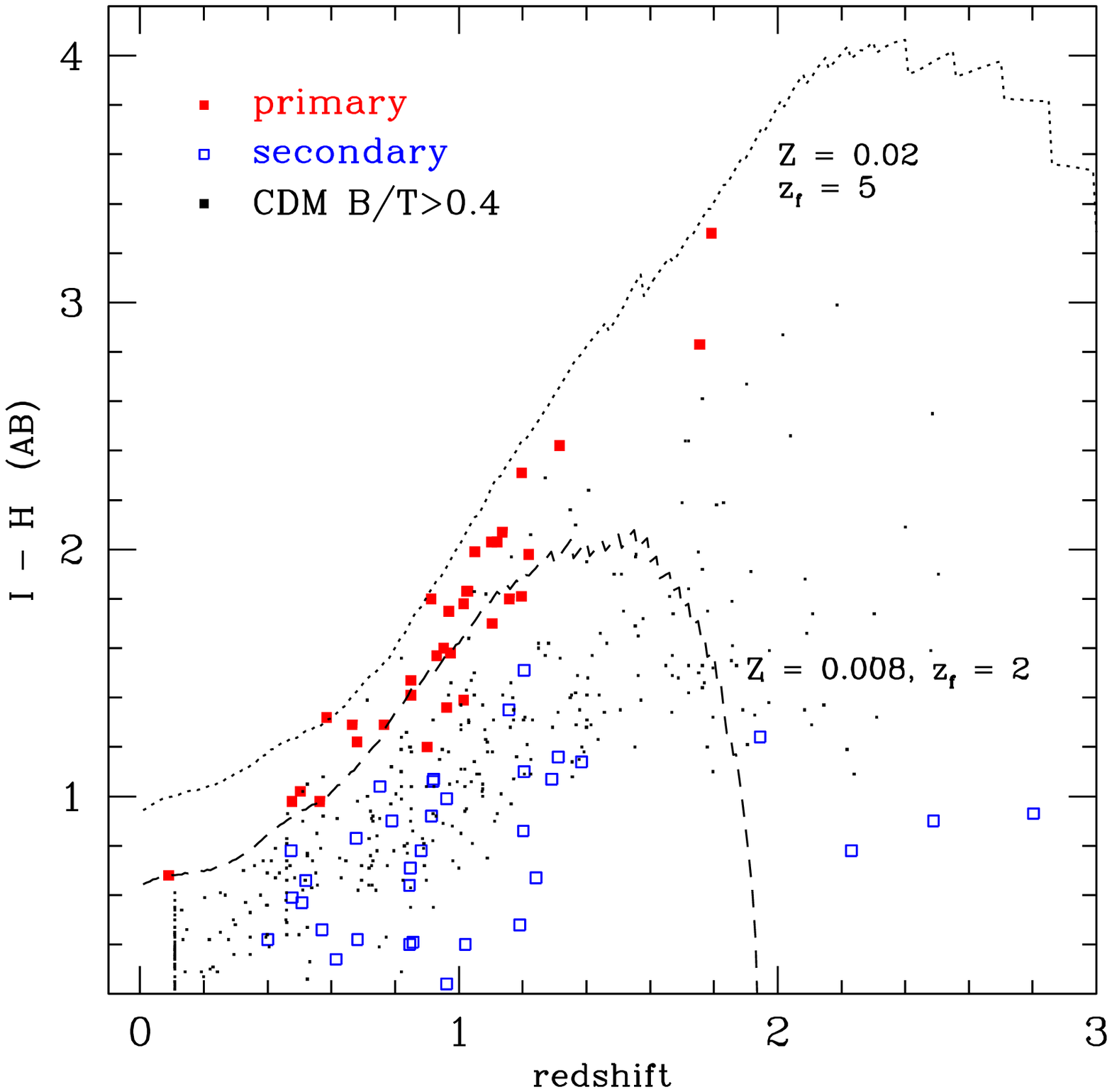}
\caption{The $I-H$ color against redshift for the primary and
secondary early-types in the HNM along with the simulated galaxies
selected from the hierarchical model with a ratio of bulge/total light
$> 0.4$ in the $\H160$ band.  Two Bruzual \& Charlot PLE models are shown for reference. }
\label{h24_eso_cdm_imhvz}
\end{figure}

It is also of interest to examine the spectral evolution of the
hierarchical model (\citet{som01}; see Section~5.3) galaxies relative
to the early-types in the HDF-N by examining the observed colors vs
redshift.  We do so in Figure~\ref{h24_eso_cdm_imhvz} which shows the
$I-H$ color against redshift.  Both the primary and secondary
early-type samples are shown, along with model early-type galaxies
selcted from the hierarchical simulations by their bulge to total
ratio.  For reference we have plotted the tracks corresponding to two
extremes of the BC PLE models, one with a solar metallicity/high $z_f$
and the other with low metallicity/low $z_f$.  As can be seen, the
hierarchical model galaxies better follow the distribution of the
secondary set of real early-type galaxies with the bluer colors at all
redshifts, but do not reproduce the red envelope of the primary
subsample.  Also the model galaxies from the simulations are far more
numerous at high redshifts compared to all the early-types that we
selected from the HDF-N.  Thus Figure~\ref{h24_eso_cdm_imhvz} does not
lend support to the hierarchical model being an adequate description
of the early-type samples.

\begin{figure}
\figurenum{15}
\plotone{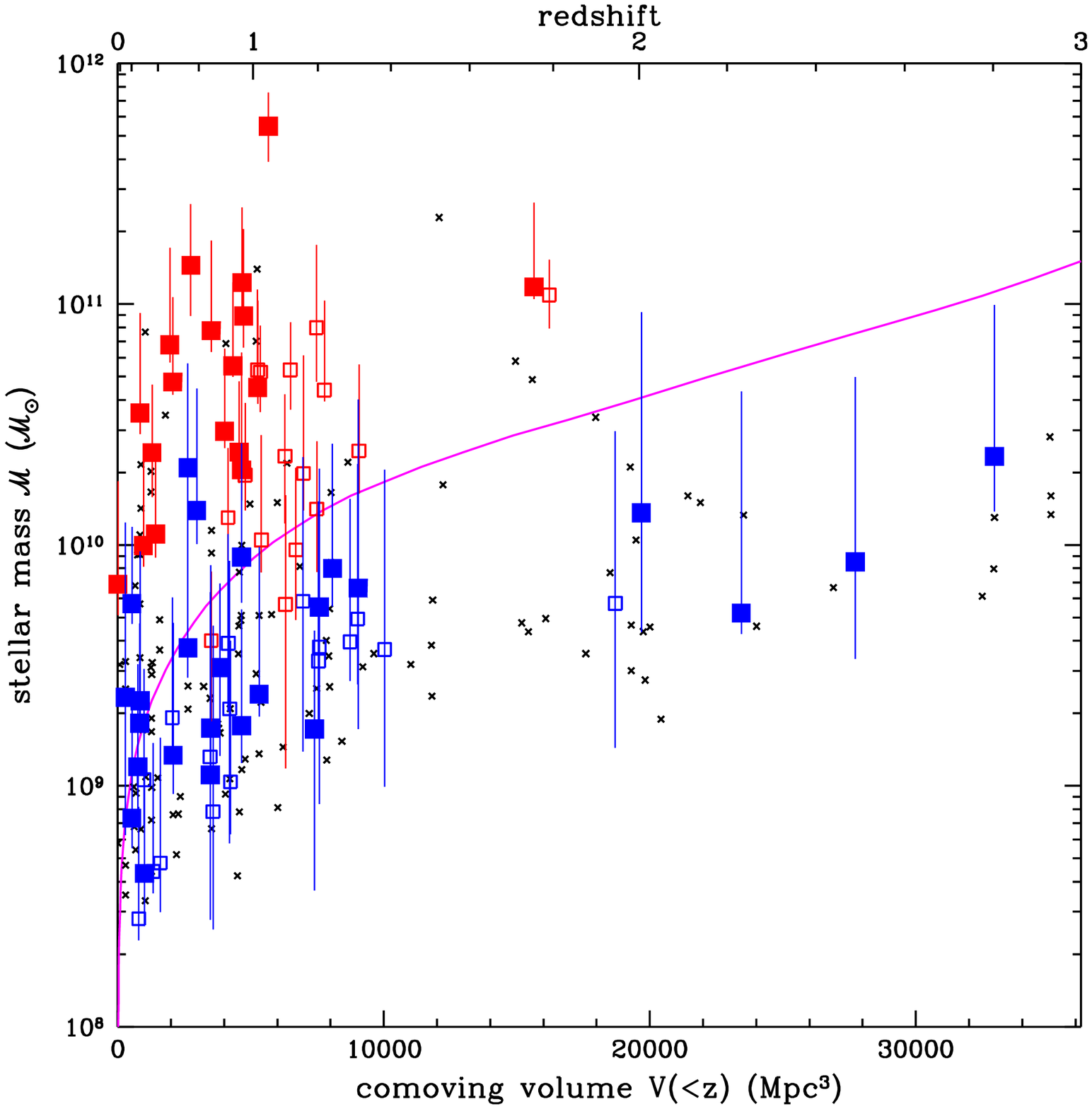}
\caption{Estimated stellar mass against comoving volume and redshift
for all galaxies in the HDF-N with $\H160 < 24.0$.  The primary sample
are plotted in red and the secondary sample are in blue.  Open squares
indicate photometric redshifts and solid squares are spectroscopic
redshifts. The error bars show the 68\% confidence range on mass,
where the upper error bar comes from the one~$\sigma$ upper limit from
the 2-component models, and the lower error bar shows the one~$\sigma$
lower limit from the 1-component model (see Papovich, Dickinson, \&
Ferguson 2001 for details).  The small black crosses are the non
early-type galaxies. The magenta line shows the mass limit
corresponding to $\H160 = 24.0$ for a PLE star burst formed at $z =
\infty$.  }
\label{mass}
\end{figure}

\citet{pap01} and \citet{dick03} report estimates
for the stellar masses of galaxies selected from the same HDF NICMOS
data. We make use of their 1-component fits which use
monotonically-declining exponential star formation rates including
dust, and show in Figure~\ref{mass} the stellar masses for the
galaxies with $\H160 < 24.0$ in the HDF-N.  Clearly the primary
early-type galaxies are more massive on average, at all redshifts,
than the secondaries.

\subsection{Scaling Relations}

\begin{figure}
\figurenum{16}
\plotone{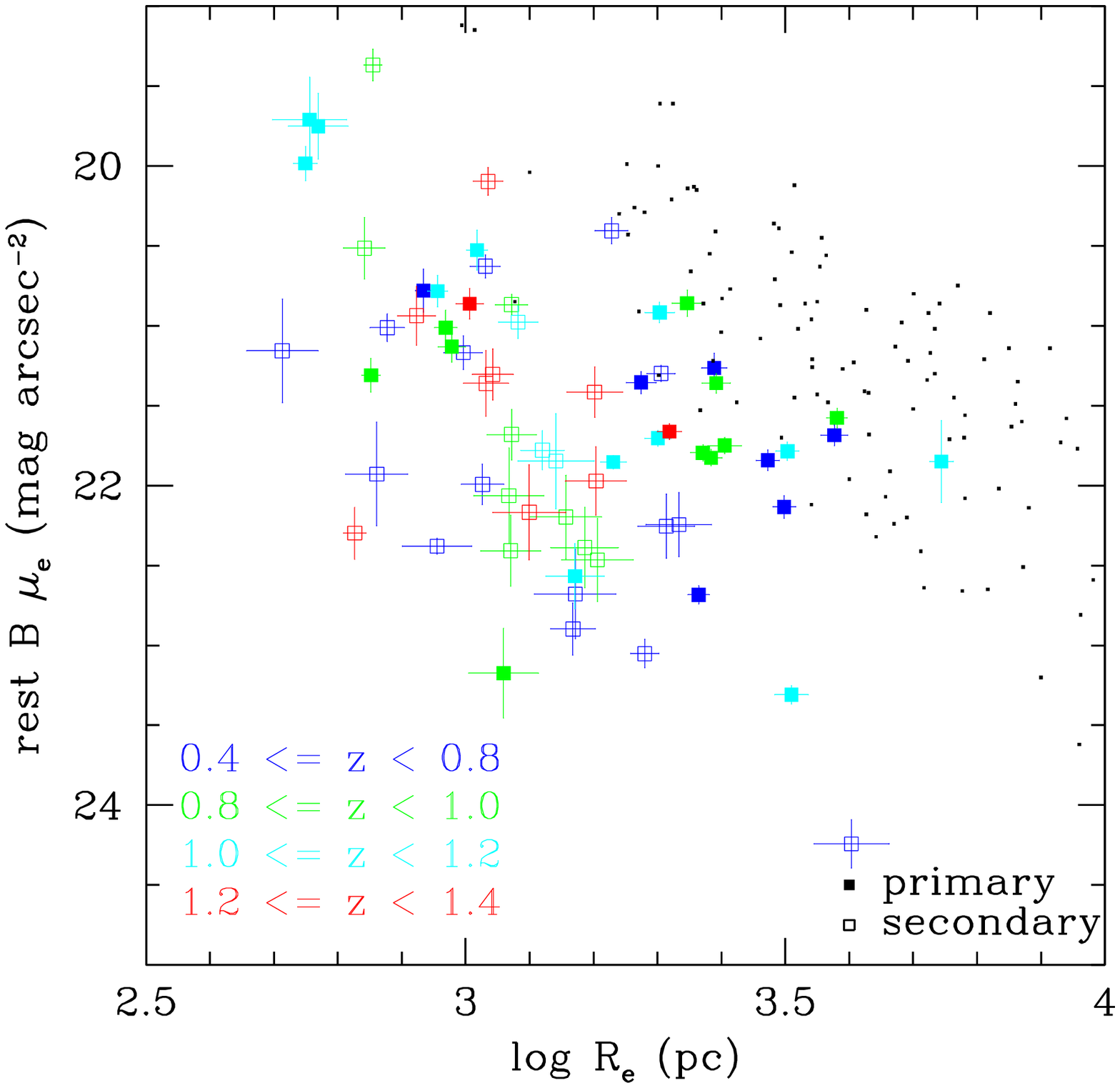}
\caption{The rest frame $B$ $\mu_e$ against log of the $R_e$ for the
two early-type samples.  One $\sigma$ errorbars are shown on each
point.  Redshift information is encoded by the color of the points as
indicated in the legend.  The values of $R_e$ and $\mu_e$ have been
corrected for the bias seen in Figure~\ref{resim}.  The solid black
dots represent ellipticals at $z \sim 0$ from \citet{jorg96}. }
\label{h24_eso_korm}
\end{figure}

Lacking velocity dispersions, we can use projections of the
Fundamental plane to examine the differences between the primary and
secondary samples of early-types.  Figure~\ref{h24_eso_korm} plots
$\mu_e$ in the rest frame $B$ against the log of the $R_e$ which were
measured by the $r^{1/4}$ fitting.  It is interesting that in
Figure~\ref{h24_eso_korm} the secondary sample of bluer galaxies is
more compact for a given surface brightness, consistent with these
galaxies having an enhancement of light in their cores, possibly due
to recent star formation \citep{men01}.  Figure~\ref{h24_stvtt} shows
that there are very few galaxies with good de Vaucouleur profiles that
are also blue and classified as morphological late types.

\begin{figure}
\figurenum{17}
\plotone{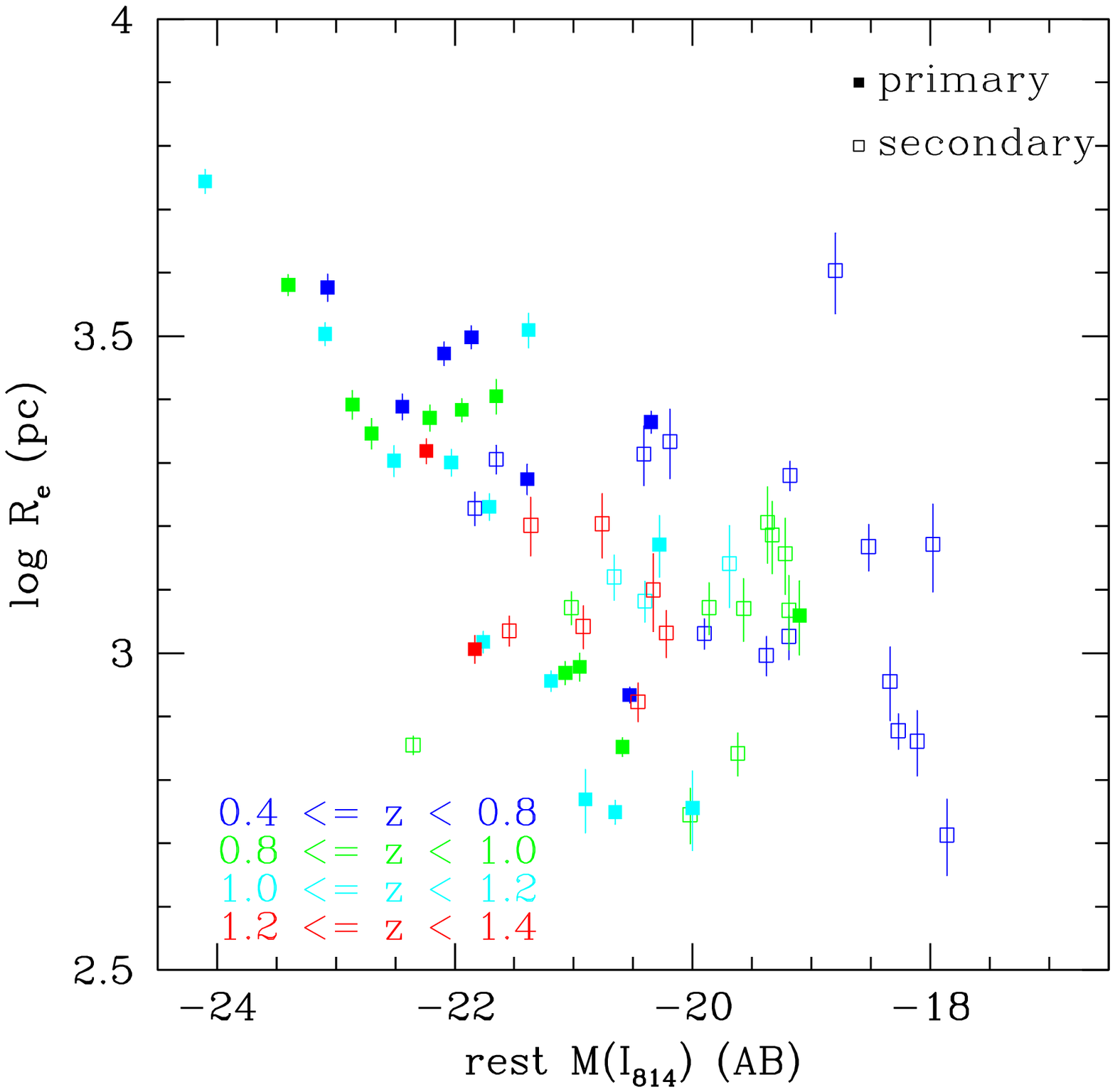}
\caption{Log of the $R_e$ against $M(\I814)$ for the
two early-type samples.  One $\sigma$ errorbars are shown on each
point based on simulations of the $r_e$ measurements.  The point color
encodes the redshift as indicated in the legend.  The values of $R_e$ have been
corrected for the bias in measuring $r_e$ found in our simulations and
seen in Figure~\ref{resim}. }
\label{h24_eso_revmi}
\end{figure}

At $z < 1$, \citet{schade99} have examined projections of the
Fundamental Plane using WFPC2 images of field elliptical galaxies
selected from the CFRS/LDSS.  They found evolution with redshift in
the relation between $R_e$ and $M(B)$ in the sense that the higher
redshift galaxies are more luminous for a given size.  The
distributions in Figure~\ref{h24_eso_revmi} agree with this trend although
the scatter is large in our measurements.  It is clear that there are
almost no large secondary early-types, whereas there are quite a few
large primary early-types: there are 4 secondaries and 17 primaries
with $R_e > 2$ kpc.

\subsection{Number Density}

\begin{figure}
\figurenum{18}
\plotone{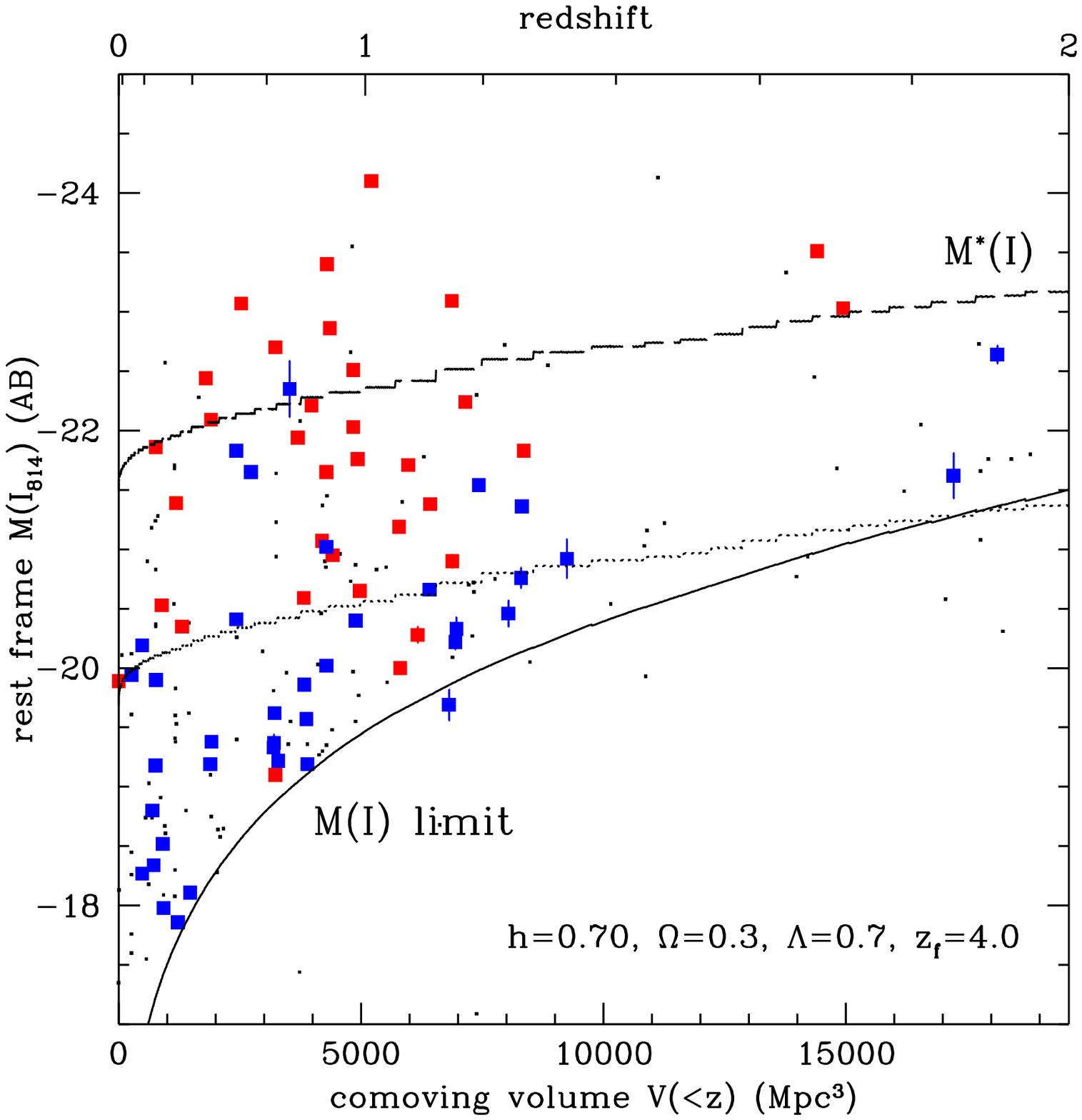}
\caption{The rest frame $M(I_{814})$ against the comoving volume of
the whole survey for the two early-type samples: the red, solid points are
the primary sample and blue, open points are the secondary sample.
Non-early-type galaxies are shown by small dots.  
Spectroscopic redshifts are used when available.   The
limiting absolute magnitude of the $H<24.0$ sample is shown by a solid
curve.  The predicted change in $M(I_{814})$ for an $L^*$ galaxy is
represented by the dashed line, calculated using a BC model with a 0.1
Gyr burst of $Z_\odot$ stars formed at $z_f = 4.0$ followed by PLE.
The dotted line is $\sim$2 magnitudes less luminous than the predicted
$M^*(\I814)$.  }
\label{h24_eso_mivv}
\end{figure}

Before making detailed comparisons with models, it is worth examining
the redshift distribution of early-types
alone. Figure~\ref{h24_eso_mivv} shows a graphical depiction of the
change in number density with redshift in the galaxies in the HDF-N
with $\H160 < 24.0$.  We have calculated the rest frame $\I814$
absolute magnitudes for all objects in the primary and secondary
samples up to $z \sim 1.8$.  At redshifts less than this limit, the
$M(I_{814})$ can be estimated by {\it interpolation} from among the
available HDF-N photometry.  Because only a few of the early-type
galaxies are at $z > 1.8$, this restriction is not unduly detrimental.
The $M(I_{814})$ measurement is plotted against the comoving volume of
the HDF-N out to the object's redshift, $V(<z)$, in
Figure~\ref{h24_eso_mivv}.  The limiting absolute magnitude of the
$\H160 < 24.0$ sample is shown by a solid curve, assuming a PLE model
with a single burst at $z = \infty$.  The predicted change in
$M(I_{814})$ for an $L^*$ galaxy is represented by the dashed line,
calculated using a BC model with a 0.1 Gyr burst of $Z_\odot$ stars
formed at $z_f = 4.0$ followed by PLE, which will be referred to below
as our standard PLE model.  The present epoch luminosity of an $L^*$
galaxy is taken from \citet{blanton} and we use our standard PLE model
to convert from their $i$-band value to the required $I_{814}$.  The
dotted line in the figure is two magnitudes less luminous than the PLE
model value for $M^*(I_{814})$, and shows the limit to which the
sample is complete up to $z \sim 2$.  A sample selected above the
dotted line should be complete out to $z = 1.8$, and would be complete
for galaxies whose stellar masses are greater than that of a
present--day, $M^\ast + 2$ elliptical galaxy.
Figure~\ref{h24_eso_mivv} demonstrates a steep decline in the space
density of the combined early-type samples, after being restricted in
$M(I_{814})$ so as to cover the same luminosity range, and after
correcting for the effects of PLE, at all redshifts up to $z \sim 2$.
This decline is also present in the number density of all galaxies at
similar redshifts $z > 1.3$.

Another way of assessing the evolution of a galaxy population is to
compare the observed number of a certain galaxy type as a function of
redshift with the predictions of a model.  The way that the early-type
galaxies in the HDF-N were counted for this comparison was more
complicated than in other analyses for the following reasons.
Whenever we select subsamples with boundaries in some parameter (like
color or T-type), the results depend strongly on the exact placement
of the boundary.  It is clear in Figure~\ref{h24_stvtt} that the exact
placement of the T-type boundaries is somewhat arbitrary and strongly
affects the numbers of galaxies which are defined to be early-types.
Its also worth recalling at this point that in some cases, the
uncertainties in the assigned T-types are fairly large.  So we tried
the following solution.  For each morphologically-selected early-type
galaxy, a Gaussian distribution in T-type is calculated with the mean and
sigma of the distribution set to be the average T-type and $\sigma(TT)$ of that
object.  Then a sum over the distribution is performed, with the limits in
T-type set by the chosen definition of an early-type, e.g.\ $-7 < TT < -2$ for
the primary sample.  This yields a weight for each object between 0 and 1
which describes its contribution to the number of early-types being
selected.  This procedure allows some contribution from objects which
have average T-types outside of the defined early-type range.  To
create a redshift histogram of early-types, the weights are added up
in each redshift bin.

\begin{figure}
\figurenum{19}
\plotone{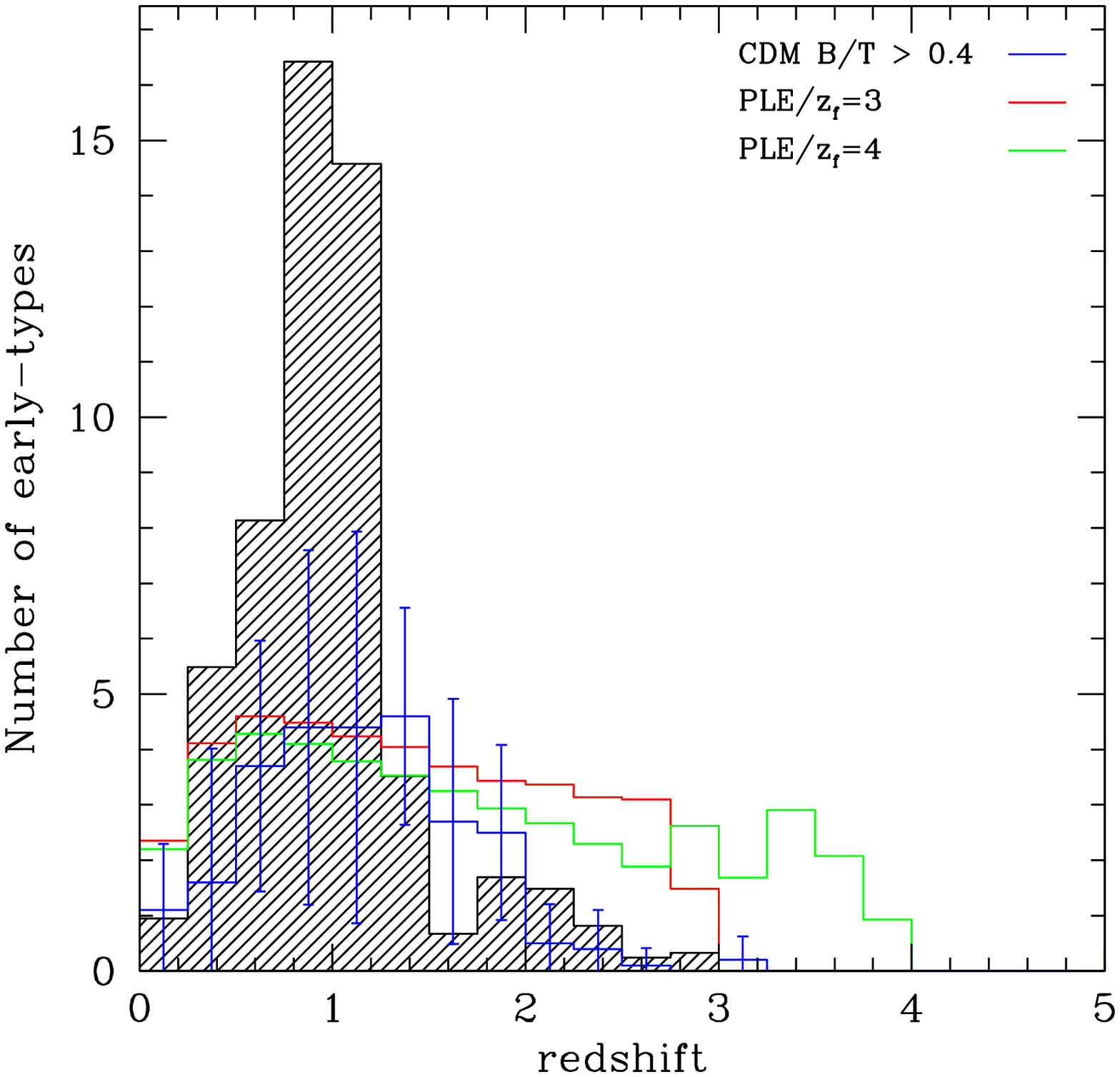}
\caption{The number of early-types against redshift in the HNM at
$\H160 < 24.0$ is shown by the shaded histogram.  Two PLE model
predictions are plotted based on the Marzke et al.\ local luminosity
function: the red histogram has $z_f = 3$ and the green histogram has
$z_f = 4$.  The PLE models assume all stars are formed in a 0.1 Gyr
burst with solar metallicity at the given $z_f$.  The blue histogram
shows a hierarchical merging prediction, along with scatterbars which
show the one $\sigma$ variation to be expected due to large scale
structure.  The hierarchical merging model is described in Somerville
et al.\ 2001.  All model calculations assume $\Lambda=0.7,
\Omega=0.3$, and $h=0.70$.}
\label{h24_eso_nvz}
\end{figure}

We have used the software described in \citet{gardner} to calculate
the number of early-type galaxies which should be found in the HNM
samples.  These calculations primarily depend on a local luminosity
function, the area and depth of the HNM sample, a galaxy evolution
model, and a chosen cosmology.  We use our standard Bruzual \& Charlot
PLE model to characterize the evolution of the early-type galaxies.
We adopt the local luminosity function of early-type galaxies from
\citet{marzke98}. For early-type galaxies only, the relevant Schechter
function parameters are $M_B^* = -20.15$, $\alpha = -1.0$, and $\phi^*
= 1.5 \times 10^{-3}$ Mpc$^{-3}$.  Clearly there are a large number of
possible models which could be generated from the \citet{gardner}
software to compare with the HNM primary and secondary samples,
depending on e.g.\ the galaxy formation epoch and the cosmological
parameters.  Let us consider two simple examples with $z_f =3$ and
$z_f=4$ in our default cosmology.  These are shown in
Figure~\ref{h24_eso_nvz} compared with the early-types that we have
selected from the HNM.  Neither of the two models based on the Marzke
et al.\ local luminosity function provides a good description of the
overall early-type redshift distribution.  In particular, the observed
early-type distribution has fewer galaxies at $z > 1.5$ than the
predictions.

To see how well the redshift distribution of the early-types could be
matched by a hierarchical galaxy formation model, we used a simulation
similar to that described in \citet{som01} and \citet{firth02}.  These
simulations include modeling of the hierarchical build-up of
structure, gas cooling, star formation and supernova feedback, galaxy
mergers, chemical evolution, stellar populations and dust. Unlike the
models presented in \citet{som01}, here we have used the
multi-metallicity stellar population models of Bruzual and Charlot
(1998 version).  Ten realizations were created and a single average
mock object catalog was extracted from all the realizations so as to
match the real $H < 24.0$ catalog obtained from the HNM in terms of
depth and area.  From the mock catalog, we selected early-type
galaxies by virtue of their having a bulge to total luminosity in the
observed $\H160$ band greater than 0.4.  While this selector appears
to be a reasonable way to choose real early-type galaxies
\citep{im02}, it is unclear how it relates to the visual method
employed in selecting the primary and secondary early-type samples
from the real data.  The redshift distribution of the early-types
morphologically selected from the hierarchical model is shown in
Figure~\ref{h24_eso_nvz} by the blue histogram.  The hierarchical
prediction is a better match to the data at $z > 1.5$ than the PLE
models, but it also underpredicts the observed number of early-types
at $z \sim 1$. \citet{firth02} found the similar result that the
hierarchical models underpredict the number density of extremely red
galaxies, which are believed to be early type galaxies at $z\sim1$,
compared to a larger-area ground-based sample selected at $H\la20$.
The errorbars on the hierarchical model histogram give an indication
of the scatter due to Poisson variation in the number of halos from
realization to realization.  The errorbars were determined by simply
calculating the root mean square in the number of early-type galaxies
at each redshift bin among the 10 realizations.  They underestimate
the expected scatter due to large scale structure, as the strong
clustering of massive halos has not been accounted for.

\begin{figure}
\figurenum{20}
\plotone{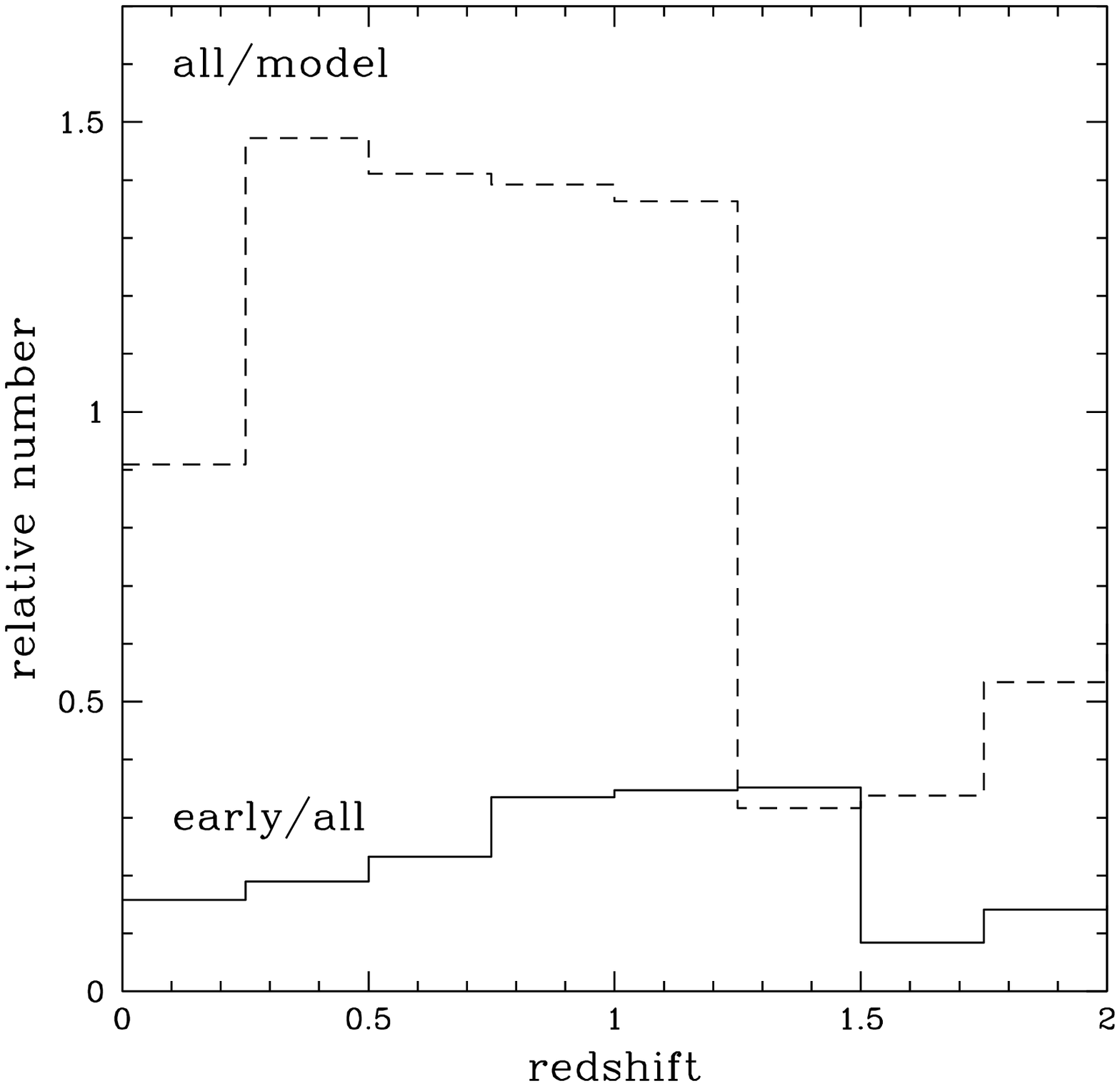}
\caption{The relative numbers of all early-types against all galaxies
at $\H160 < 24.0$ is shown by the solid line, and of all galaxies
against the prediction of the CDM model for all galaxies in the
$\H160 < 24.0$ sample.}
\label{h24_fracs_vs_z}
\end{figure}

The relative redshift distributions of the combined early-type
samples, all of the galaxies at $\H160 < 24.0$, and all of the
galaxies predicted by the CDM model to be at $\H160 < 24.0$ in the
HDF-N are compared in Figure~\ref{h24_fracs_vs_z}.  This plot shows
that the fraction of early-type galaxies in the total $\H160 < 24.0$
sample changes by only a small amount up to $z \sim 2$.  But the
comparison of the entire $\H160 < 24.0$ sample with all the galaxies
in the hierarchical model indicates large changes in this ratio as a
function of redshift.

\begin{figure}
\figurenum{21}
\plotone{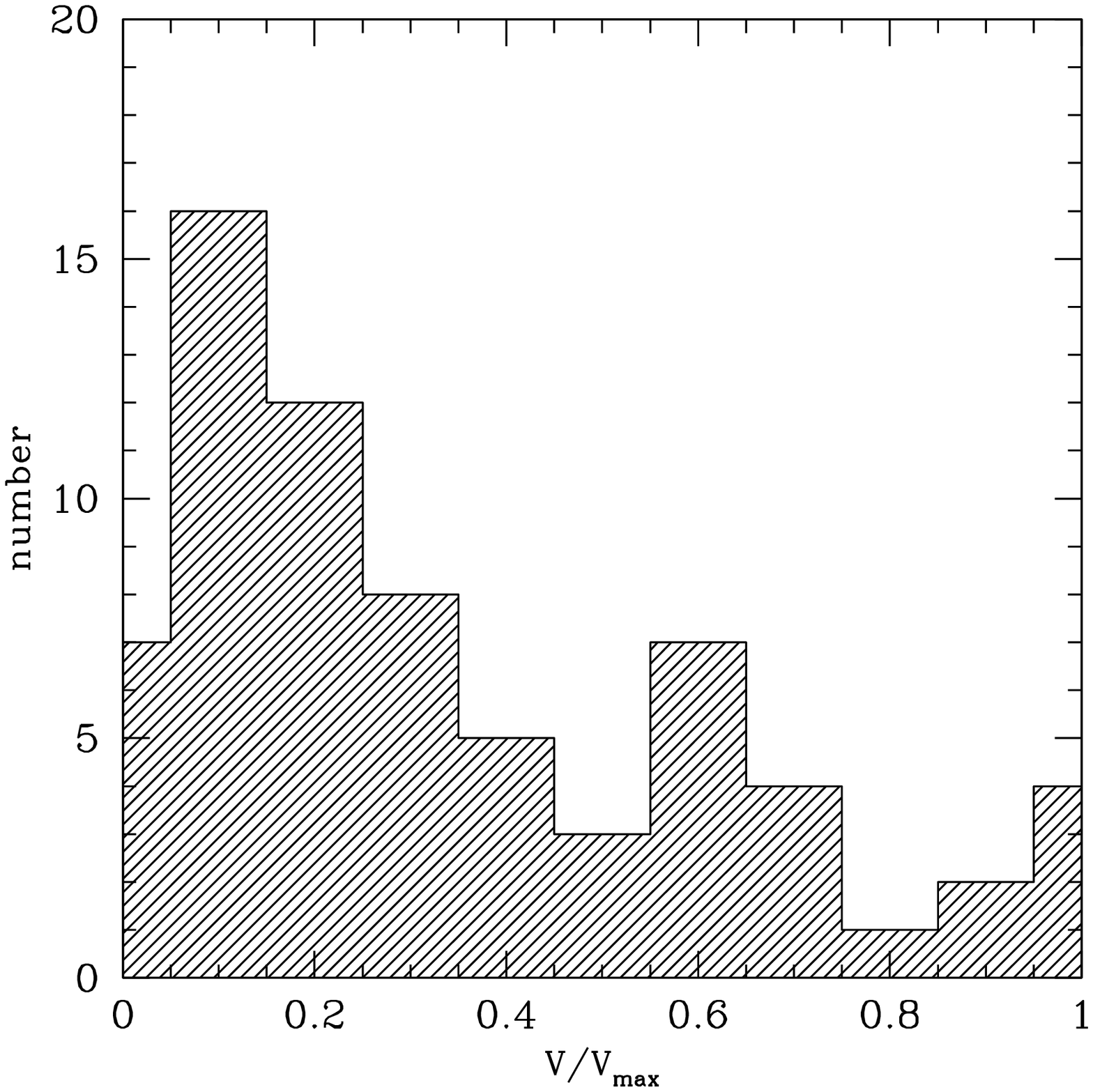}
\caption{A $V/V_{max}$ test for all early-types in the HNM sample at $H
< 24$.} 
\label{h24_eso_vvmax}
\end{figure}

Another approach to examining the space density evolution is to use
the $\langle V/V_{max} \rangle$ statistic \citep{schmidt68}.  This ratio
is a measure of the position of a galaxy within the observable volume.
In the case of constant space density, the individual values of the
ratio will be uniformly distributed between 0 and 1, i.e. $\langle
V/V_{max} \rangle = 0.5$.  The highest possible $z$ for each galaxy is
calculated such that it would still be in the $\H160 < 24.0$ sample.
The volume for this maximum redshift is $V_{max}$.  The actual
redshift of the galaxy gives the actual volume $V$.  We made the
$V_{max}$ calculation using our standard PLE model (making the galaxy
brighter in the past) to estimate the upper redshift limit for a given
galaxy.  The result is shown in Figure~\ref{h24_eso_vvmax} from which
we calculate that $\langle V/V_{max} \rangle = 0.31 \pm 0.04$ for the
combined primary+secondary samples.  If only the primary sample was
used, the resulting $\langle V/V_{max} \rangle = 0.22 \pm 0.05$.  If
we had assumed no-evolution in calculating $V_{max}$ then $\langle
V/V_{max} \rangle = 0.37 \pm 0.04$ for all early-types.

Our analysis confirms previous findings (some of which based on the
HDF-N) to the effect that the space density of early-type galaxies is
substantially smaller at $z \gtrsim 1.4$ than at lower redshift
\citep{zepf97,franc98,barg99,im99,men99,abraham99,treu99}.  In the
case of the HDF-N field, it is likely that no reasonable model can fit
in detail the redshift distribution of not just the early-type but all
galaxies due to the way that a large scale structure distorts the
redshift distribution at $z \sim 1$.  There are two well-known
redshift spikes in the HDF-N galaxy distribution at $z = 0.96$ and $z
= 1.02$ and both of these spikes are rich in early-type galaxies.

\subsection{The Highest Redshift Early-type Galaxies?}

While one of the main results of our study is the relative paucity of
$z > 1$ early-type galaxies in the HDF-N, there are two objects in our
HNM catalog apparently at redshifts approaching $z \sim 2$ that seem
to qualify as massive elliptical galaxies.  Object \# 731 is the host
galaxy of SN 1997ff \citep{riess01} and has a $z_{phot} = 1.65$.  A
possible spectroscopic redshift of $z = 1.755$ has been obtained as
well \citep{riess01}.  This object was classified as an elliptical by
our visual inspection of its NIC3 image with an average T-type of
$-3.3$, although with a significant spread in the assigned TT, and the
surface brightness profile fitting gave a good fit to a de Vaucouleur
law.  In the rest frame, the galaxy is as red ($B - V = 0.6$) and
luminous ($M_I = -23.5$) at its redshift as expected for a massive,
fully formed elliptical galaxy.

The second object which appears to be a high redshift early-type
galaxy is \#882 in our catalog with $z_{phot} = 1.79$.  It is quite
close to \#731 on the sky but not as well known and has not been the
target of spectroscopy to determine its redshift.  However, compared
to \#731, \#882 has an even redder color and is similarly luminous in
the rest frame $M(I_{814})$ band.  Although the assigned T-type of
$-3.3$ indicates that this object appears to be an elliptical based on
its morphology in our NIC3 image, the profile fitting gave only a fair
fit to a de Vaucouleur law.  For both objects the SEDs indicate
relatively little recent star formation, and the fitting in
\citet{dick03} gave masses of $\sim 10 \times 10^{11} M_\odot$ which
is very large given the high redshifts.  Both of these objects are in
our primary early-type sample.

\section{Discussion}

We have made a map of the HDF-N in the F110W and F160W filters
using Camera 3 of the NICMOS onboard $HST$.  Using these data, an
object sample was selected from the F160W mosaic, which is $\sim$90\%
complete at $\H160_{AB} \sim 26.0$; here we reported results on the
sample limited to $\H160_{AB} < 24.0$.  Hubble types were determined
by visual inspection of the \H160 images.  Galaxy spectral types were
obtained from the procedure of estimating photometric redshifts.  de
Vaucouleur law profiles were fit to the \H160 surface brightness
profiles and categorized by the goodness of the fit.

We have selected two samples of early-type galaxies.  The primary
sample consists of 34 galaxies with $-7 < $ T-type $ < -2$ and
early-type galaxy spectral types.  All these objects have fair or good
$r^{1/4}$ law profile fits.  The secondary sample of 42 galaxies has
$-7 < $ T-type $ < 0$, later-type galaxy spectral types, and fair or
good $r^{1/4}$ law fits.  The primary sample is more luminous (its
average rest frame magnitude $M_B = -20.27$ vs $M_B = -19.4$ for the
secondary sample) and generally physically larger at the same $M_B$
compared to the secondary one, independent of redshift.  There are
no major differences in the $\H160$ morphologies of the primary and secondary
samples, and there is no difference in the nearby environments of the
primary vs secondary samples of early-types, hence there is no
evidence that the secondary sample is preferentially undergoing galaxy
interactions.  However the secondary sample is considerably bluer
(rest frame $B-V = 0.42$ on average vs $B-V = 0.71$ for the primary
sample), indicating that diffuse, widespread star formation has
recently occurred in the secondary early-types.  This would agree with
the results of \citet{trager00} which show that $z=0$ ellipticals
have a wide range of ages, from 1.5 Gyr up to a Hubble time.  Also,
the galaxies identified in our secondary sample may be similar to the
ellipticals in the HDF which been shown to have blue cores and appear
to be forming stars \citep{men01}.

The relationship between the primary and secondary samples does not
appear to be direct.  Because they are blue and less luminous
in the rest frame optical, the secondary galaxies are unlikely to
evolve into the primary galaxies since they are less massive than the
primary sample galaxies.  After their current episodes of star
formation end, the secondary galaxies will become redder but less
luminous and their stellar masses will not increase.  The secondaries
could be the building blocks from which the primary galaxies are made
through mergers.  However their redshift distributions are broadly
similar; one would expect there to be more secondaries at higher
redshift in order to make the primaries that we see at $z < 1.5$ in
the HDF-N.  

The redshift distribution of the early-type samples was examined to
test the predictions of monolithic and hierarchical formation
scenarios.  Both the primary and secondary samples largely disappear
at $z > 1.4$; there are only a few early-types from the secondary
sample up to $z \sim 2.5$.  The observed redshift distribution does
not match that predicted by a monolithic scenario.  For a cosmology of
$h=0.7$, $\Omega = 0.3$, $\Lambda = 0.7$, the predicted redshift
distribution of passively evolving early-types formed at high redshift
shows a deficit at $z \sim 1$ and an overabundance at $z > 1.5$ with
respect to the primary and to the primary+secondary samples in the
HNM.  A $V/V_{max}$ test agrees with this result.  When the effects of
passive luminosity evolution are included in the calculation, the mean
value of $V_{max}$ for the primary sample is $0.22 \pm 0.05$, and
$0.31 \pm 0.04$ for the combined primary+secondary sample.  A
hierarchical formation model such as that of \citet{kauf96} or
\citet{som01} better matches the overall redshift distribution of the
early-types, with the exception of the spike at $z \sim 1$, though it
still overpredicts the number at $z > 1.5$. 

Our results may be affected by several forms of bias.  First, the
apparently small number of early-types at $z > 1$ could be due to a
selection bias, if for example high redshift ellipticals are heavily
obscured by dust during their formative stage.  The SCUBA detections
of high redshift galaxies indicates that this may be a real
possibility, although the connection between the SCUBA sources and
elliptical galaxies is still unknown.  A more complicated situation
would result if the progenitors of ellipticals galaxies are present at
high redshift but we fail to select them because they are
morphologically different from a present epoch elliptical.  This could
happen simply because the elliptical was undergoing a merger at the
time of observation, or because there really is a morphological
transformation from late-type to early-type which is the origin of
present-epoch ellipticals.  However, if ellipticals are morphologically
different preferentially at $z > 1$ and if they all form at the same
time they should all (or nearly all) look morphologically similar at
lower redshifts.  The fact that we do not find the same density of
ellipticals is a sign that there are multiple formation mechanisms
and/or times for ellipticals.  The fact that all nearby ellipticals
have old stellar pops is a sign that the stars in ellipticals must
exist at $z \sim 1.5$, but are not in what we can identify as ellipticals
at that time.  This would be an argument against the idea that
monolithic collapse for some ellipticals happens later, at e.g.\ $z
\sim 1$.

Previous studies of the N(z) of red and/or early-type galaxies have
reported mixed results as to the number of such galaxies at $z > 1$,
as summarized in the Introduction.  Field galaxy studies that use
morphological identification of early-types tend to show a relatively
small number at $z > 1$ \citep{treu99,ben99}, while the larger area
surveys that use only colors to identify early-types find a nearly
constant space density \citep{cim00,pat01}.  Our results on N(z) in
the particular case of the HDF-N agree with previous studies
\citep{zepf97,franc98,barg99} in finding few
morphologically-identified early-types at $z > 1.5$.  It is also true
and perhaps significant that we find that there are not even many blue
spheroidals at these higher redshifts, although there are a few at $2
< z < 3$.  As for the complete disappearance of the red galaxies in
the primary sample at $z > 2$, the cause may be real morphological
evolution which results in the higher redshift galaxies being left out
of our morphological definition of an early-type sample \citep{con03}.

The HDF-N is one very small window on the universe.  The N(z) of the
early-type galaxies (indeed all galaxies in our sample) appears to be
dominated by large scale structure at $z > 0.8$; it seems impossible
for any reasonable model to completely describe the observed N(z) of
the HDF-N.  Clearly, data similar to those employed here in terms of
broad wavelength coverage, high resolution, and depth, but covering
much larger areas are necessary to make further progress on the
questions examined in this paper.  The data from campaigns such as
GOODS that are planned to probe the $z > 1$ era over wider areas using
e.g.\ ACS on HST, SIRTF, and DEIMOS at Keck and VIRMOS at the VLT
should enable significantly better understanding of the origin and
evolution of early-type field galaxies.

\acknowledgements

We thank the STScI staff who helped to ensure that the NICMOS
observations were carried out in an optimal manner.  Support for this
work was provided by NASA through grant GO-07817 from the Space
Telescope Science Institute, which is operated by the Association of
Universities for Research in Astronomy, Inc., under NASA contract
NAS5-26555. Part of this work was performed
under the auspices of the U.S.\  Department of Energy by University of
California, Lawrence Livermore National Laboratory under contract
No.\ W-7405-Eng-48.

\begin{figure}
\figurenum{7b}
\plotone{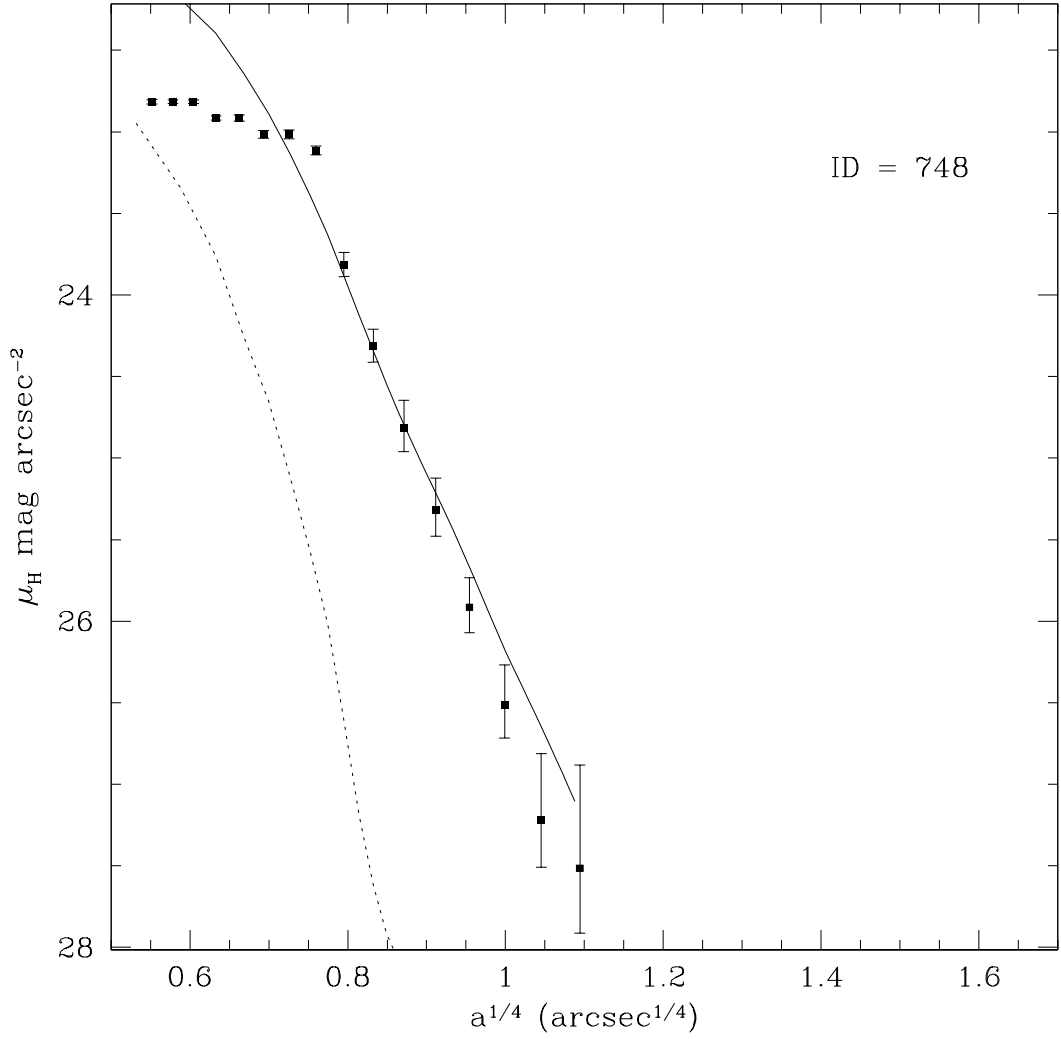}
\caption{An example of a fair fit.}
\end{figure}

\begin{figure}
\figurenum{7c}
\plotone{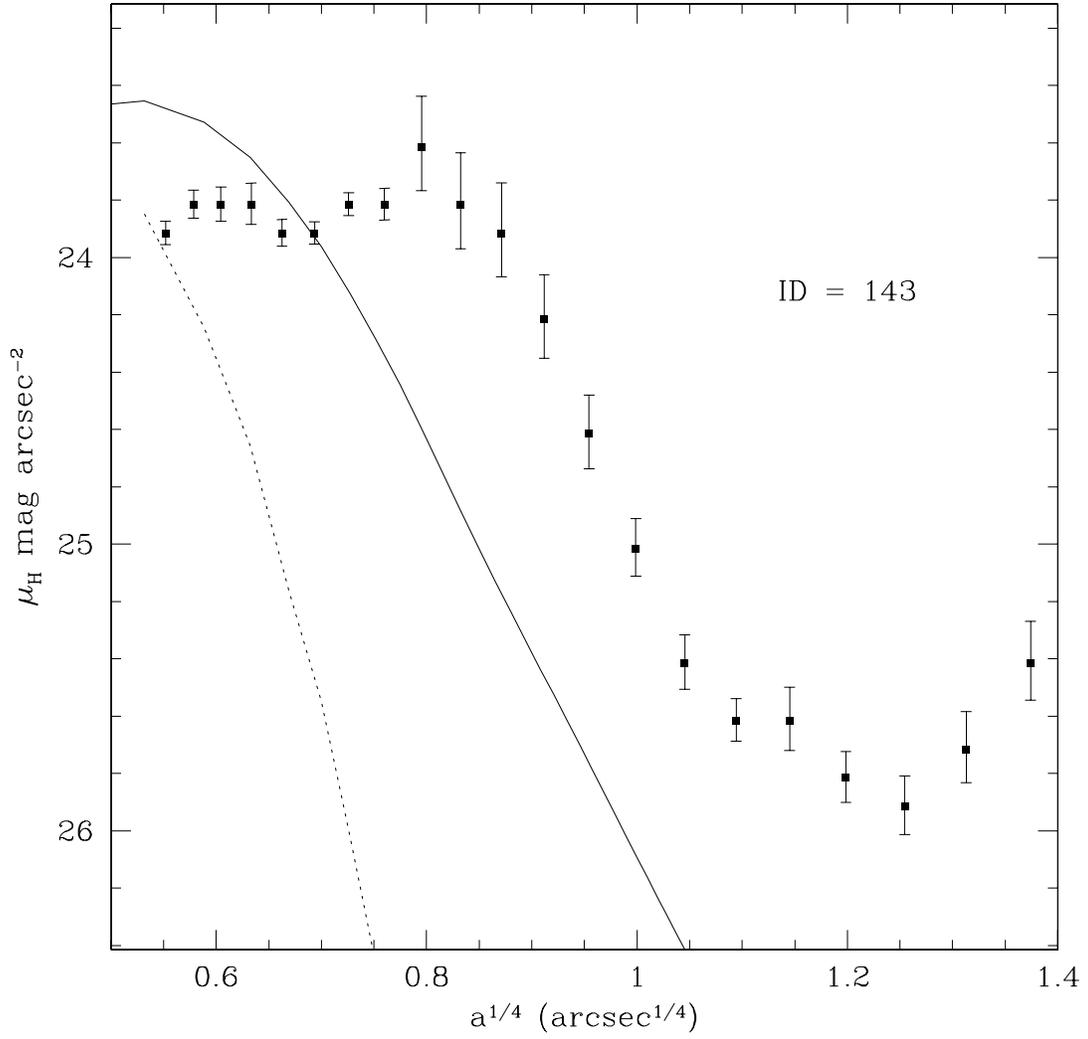}
\caption{An example of a poor fit.}
\end{figure}

\begin{figure}
\figurenum{10}
\caption{Color composite images of the primary early-type sample.  For
each galaxy, $IJH$ images of a 10 arcsec area are shown on the right
and $BVI$ images on the left.  The redshifts are shown, with a p in
parentheses to indicate the objects for which no spectroscopic
redshift is available.  The galaxy ID number is shown on the left. }
\label{primon1}
\end{figure}

\begin{figure}
\figurenum{10b}
\caption{continued}
\end{figure}

\begin{figure}
\figurenum{11}
\caption{Color composite images of the secondary early-type sample.
For each galaxy, $IJH$ images of a 10 arcsec area are shown on the
right and $BVI$ images on the left.The redshifts are shown, with a p
in parentheses to indicate the objects for which no spectroscopic
redshift is available.  The galaxy ID number is shown on the left. }
\label{secmon1}
\end{figure}

\begin{figure}
\figurenum{11b}
\caption{continued}
\end{figure}

\clearpage

\clearpage

\begin{deluxetable}{lrrlllllrclclrc}
\tabletypesize{\footnotesize}
\rotate
\tablecaption{HNM $\H160 < 24.0$ Sample}
\tablewidth{0pt}
\tablehead{
\colhead{ID} & 
\colhead{RA\tablenotemark{a}} &
\colhead{Dec.\tablenotemark{a}} & 
\colhead{$H_{160}^{k}$} &
\colhead{$I-H$} & 
\colhead{$z_{phot}$\tablenotemark{b}} &
\colhead{$z_{spec}$\tablenotemark{c}} & 
\colhead{ST\tablenotemark{d}} & 
\colhead{~~TT\tablenotemark{e}} & 
\colhead{$\sigma$\tablenotemark{f}} & 
\colhead{$r_e$\tablenotemark{g}} &
\colhead{$\mu_e$\tablenotemark{h}} & 
\colhead{q\tablenotemark{i}} & 
\colhead{~$R_e$\tablenotemark{j}} &
\colhead{$M_{\I814}$} \\
\colhead{} & \colhead{} & \colhead{} & \colhead{AB} &
\colhead{AB} & \colhead{} & \colhead{} & \colhead{}
& \colhead{} & \colhead{} & \colhead{\arcsec} &
\colhead{mag/\sq\arcsec} & \colhead{} & \colhead{~kpc} &
\colhead{rest AB} 
}
\startdata
   32 &12:36:46.91 &62:14:22.1  & 19.47 & 0.47 &0.140  &    0  & 0.970 & -10.00 & 0.00 & 0.17 & 19.42 &  p &  0.00 &-11.78    \\
   40 &12:36:48.63 &62:14:23.2  & 23.52 & 0.48 &0.946  &   -1  & 0.670 &  -8.00 & 2.94 & 1.37 & 27.36 &  p & 10.46 &-19.27    \\
   62 &12:36:43.55 &62:14:09.0  & 23.95 & 1.05 & 2.44  &   -1  & 0.710 &   3.00 & 3.63 & 0.11 & 21.77 &  p &  0.89 &-21.81    \\
   67 &12:36:44.07 &62:14:10.1  & 23.86 & 0.60 &2.313  &2.267  & 0.820 &   3.75 & 5.16 & 0.13 & 22.15 &  f &  1.07 &-20.88    \\
  108 &12:36:52.80 &62:14:32.1  & 20.69 & 0.18 &0.431  &    0  & 0.590 &  -9.29 & 1.89 & 0.11 & 19.47 &  p &  0.00 &-10.84    \\
  109 &12:36:48.23 &62:14:18.5  & 23.49 & 0.74 &2.244  &2.009  & 0.740 &   5.20 & 3.83 & 0.29 & 23.33 &  f &  2.43 &-21.36    \\
  110 &12:36:48.30 &62:14:16.6  & 22.52 & 0.90 &2.443  &2.005  & 0.690 &  -3.00 & 2.45 & 0.17 & 21.40 &  f &  1.42 &-22.64    \\
  116 &12:36:47.17 &62:14:14.3  & 22.30 & 1.27 &0.625  &0.609  & 0.080 &   0.00 & 4.55 & 0.25 & 20.85 &  f &  1.69 &-19.62    \\
  121 &12:36:44.85 &62:14:06.1  & 23.14 & 1.57 &  1.9  &   -1  & 0.430 &   3.20 & 1.91 & 0.23 & 22.97 &  f &  1.93 &-22.05    \\
  133 &12:36:48.09 &62:14:14.5  & 23.95 & 0.61 & 0.92  &   -1  & 0.550 &   1.50 & 5.90 & 0.05 & 20.02 &  f &  0.39 &-19.23    \\
  143 &12:36:46.48 &62:14:07.6  & 23.05 & 0.41 &0.599  & 0.13  & 0.750 &   2.40 & 5.55 & 1.99 & 27.36 &  p &  4.62 &-15.33    \\
  144 &12:36:46.34 &62:14:04.7  & 19.89 & 1.36 &0.908  &0.962  & 0.130 &  -3.14 & 2.55 & 0.59 & 21.78 &  g &  4.68 &-23.40    \\
  152 &12:36:48.34 &62:14:12.4  & 23.27 & 0.63 &0.975  &1.015  & 0.610 &   6.50 & 3.49 & 0.83 & 25.25 &  p &  6.63 &-19.97    \\
  154 &12:36:50.35 &62:14:18.7  & 22.66 & 0.49 &0.815  &0.819  & 0.570 &   2.21 & 1.82 & 1.03 & 25.25 &  p &  7.80 &-20.14    \\
  158 &12:36:51.38 &62:14:21.0  & 23.11 & 0.26 &0.438  &0.439  & 0.640 &   1.64 & 2.12 & 0.49 & 24.28 &  p &  2.79 &-18.18    \\
  161 &12:36:49.57 &62:14:14.8  & 23.61 & 0.48 &0.953  &   -1  & 0.720 &   7.79 & 2.27 & 1.09 & 26.17 &  f &  8.46 &-19.30    \\
  163 &12:36:49.82 &62:14:14.8  & 22.44 & 1.29 &1.366  & 1.98  & 0.440 &   5.14 & 2.39 & 1.67 & 25.61 &  f & 13.99 &-22.73    \\
  199 &12:36:48.55 &62:14:07.9  & 23.58 & 1.70 &1.104  &   -1  & 0.130 &  -5.00 & 3.50 & 0.05 & 19.66 &  f &  0.41 &-20.00    \\
  214 &12:36:49.50 &62:14:06.8  & 20.76 & 1.04 &0.843  &0.752  & 0.260 &  -3.71 & 1.60 & 0.25 & 20.50 &  f &  1.84 &-21.83    \\
  215 &12:36:46.03 &62:13:56.4  & 23.30 & 0.92 &0.914  &   -1  & 0.340 &  -1.42 & 2.76 & 0.17 & 22.15 &  f &  1.34 &-19.86    \\
  229 &12:36:51.33 &62:14:11.3  & 23.80 & 0.66 &2.013  &   -1  & 0.800 &   4.00 & 7.84 & 0.21 & 22.90 &  p &  1.75 &-20.31    \\
  232 &12:36:53.64 &62:14:17.7  & 23.01 & 0.47 & 0.51  &0.517  & 0.490 &   1.86 & 2.72 & 0.49 & 23.78 &  f &  3.05 &-18.67    \\
  257 &12:36:51.39 &62:14:08.2  & 23.80 & 1.51 &1.204  &   -1  & 0.260 &  -3.40 & 2.06 & 0.17 & 22.83 &  f &  1.43 &-20.33    \\
  274 &12:36:50.09 &62:14:01.1  & 23.58 & 1.04 &2.343  &2.237  & 0.680 &   4.92 & 3.61 & 0.33 & 23.96 &  g &  2.72 &-22.10    \\
  282 &12:36:53.58 &62:14:10.2  & 23.97 & 0.53 &3.432  &3.181  & 0.720 &   6.10 & 2.94 & 0.35 & 24.28 &  p &  2.65 &-20.35    \\
  289 &12:36:50.27 &62:13:59.2  & 23.69 & 0.93 &1.326  &   -1  & 0.670 &   0.70 & 3.31 & 0.23 & 23.33 &  p &  1.92 &-20.05    \\
  308 &12:36:52.85 &62:14:04.9  & 22.66 & 0.62 &0.649  &0.498  & 0.390 &   7.25 & 4.20 & 0.43 & 23.63 &  p &  2.62 &-18.91    \\
  313 &12:36:51.97 &62:14:00.9  & 22.35 & 0.71 & 0.54  &0.559  & 0.370 &   1.50 & 2.55 & 0.77 & 23.70 &  p &  4.98 &-19.41    \\
  330 &12:36:49.99 &62:13:51.0  & 22.46 & 0.64 &0.851  &0.851  & 0.880 &   7.00 & 1.00 & 1.51 & 26.17 &  p & 12.35 &-20.94    \\
  331 &12:36:49.42 &62:13:46.9  & 17.51 & 0.68 &0.093  &0.089  & 0.000 &  -4.29 & 1.25 & 0.69 & 19.42 &  f &  1.15 &-19.89    \\
  333 &12:36:47.17 &62:13:41.9  & 22.62 & 1.07 &1.268  &1.313  & 0.530 &  -0.67 & 3.39 & 0.21 & 22.17 &  g &  1.76 &-21.36    \\
  340 &12:36:51.78 &62:13:53.9  & 20.03 & 1.08 &0.541  &0.557  & 0.160 &   1.36 & 1.97 & 0.59 & 21.67 &  p &  3.81 &-21.68    \\
  356 &12:36:55.49 &62:14:02.7  & 22.24 & 0.81 &0.591  &0.564  & 0.270 &   3.14 & 1.75 & 1.13 & 24.58 &  p &  7.34 &-19.53    \\
  359 &12:36:42.03 &62:13:21.4  & 23.59 & 0.40 &0.936  &0.846  & 0.620 &  -0.80 & 4.60 & 0.25 & 23.44 &  f &  1.91 &-19.37    \\
  360 &12:36:52.74 &62:13:54.8  & 21.47 & 0.53 & 1.28  &1.355  & 0.740 &   9.60 & 0.55 & 1.95 & 25.61 &  p & 16.42 &-22.55    \\
  383 &12:36:42.37 &62:13:19.3  & 23.33 & 0.71 &0.856  &0.847  & 0.420 &  -6.00 & 3.00 & 0.07 & 20.47 &  f &  0.54 &-19.62    \\
  387 &12:36:55.58 &62:13:59.9  & 23.57 & 0.26 &0.592  &0.559  & 0.580 &   2.58 & 1.43 & 0.37 & 24.05 &  f &  2.39 &-18.30    \\
  388 &12:36:54.09 &62:13:54.4  & 21.69 & 0.80 &0.836  &0.851  & 0.390 &  -2.00 & 2.45 & 0.43 & 22.51 &  p &  3.30 &-21.23    \\
  402 &12:36:52.22 &62:13:48.1  & 23.90 & 0.46 & 0.57  &   -1  & 0.460 &  -2.60 & 3.50 & 0.07 & 21.05 &  f &  0.45 &-17.86    \\
  411 &12:36:45.41 &62:13:25.9  & 22.33 & 0.34 & 0.43  &0.441  & 0.600 &   2.00 & 2.06 & 0.31 & 22.39 &  g &  1.77 &-19.03    \\
  412 &12:36:45.86 &62:13:25.8  & 20.37 & 0.58 &0.424  &0.321  & 0.430 &   3.29 & 1.38 & 1.99 & 24.58 &  p &  9.27 &-20.12    \\
  432 &12:36:47.44 &62:13:30.1  & 23.13 & 0.49 &0.845  &   -1  & 0.940 &   0.25 & 4.51 & 0.51 & 24.42 &  p &  3.93 &-19.81    \\
  448 &12:36:55.52 &62:13:53.5  & 21.90 & 0.80 &  1.1  &1.147  & 0.600 &   4.08 & 2.68 & 1.17 & 24.76 &  p &  9.65 &-21.78    \\
  457 &12:36:54.77 &62:13:50.8  & 23.57 & 0.64 &0.845  &   -1  & 0.470 &  -0.30 & 2.86 & 0.23 & 23.13 &  f &  1.77 &-19.33    \\
  466 &12:36:52.99 &62:13:44.2  & 23.40 & 0.71 &2.024  &   -1  & 0.770 &   9.50 & 0.50 & 0.43 & 24.42 &  f &  3.58 &-21.76    \\
  477 &12:36:42.11 &62:13:10.2  & 23.04 & 0.74 &4.351  &   -1  & 0.540 &  -6.60 & 2.87 & 0.05 & 19.71 &  p &  0.00 &-23.41    \\
  488 &12:36:48.57 &62:13:28.3  & 22.34 & 0.63 & 0.85  &0.958  & 0.460 &   3.86 & 2.91 & 1.19 & 24.97 &  p &  9.44 &-20.90    \\
  503 &12:36:55.06 &62:13:47.1  & 23.94 & 0.62 &2.393  &2.233  & 0.810 &   2.00 & 5.35 & 0.09 & 21.64 &  g &  0.74 &-21.54    \\
  519 &12:36:42.52 &62:13:05.2  & 23.57 & 0.79 &1.097  &   -1  & 0.900 &   3.33 & 6.11 & 1.17 & 26.17 &  p &  9.58 &-19.97    \\
  520 &12:36:42.72 &62:13:07.1  & 21.03 & 1.23 &0.666  &0.485  & 0.050 &  -0.36 & 1.97 & 0.57 & 22.15 &  g &  3.56 &-20.50    \\
  522 &12:36:44.10 &62:13:10.8  & 22.96 & 1.07 &3.041  &2.929  & 0.550 &   7.20 & 3.56 & 0.45 & 23.50 &  p &  3.50 &-23.26    \\
  537 &12:36:42.16 &62:13:05.1  & 23.98 & 1.30 &1.726  &   -1  & 0.520 &   5.30 & 3.83 & 0.13 & 22.27 &  f &  1.10 &-20.77    \\
  539 &12:36:50.29 &62:13:29.8  & 23.74 & 1.23 &1.934  &   -1  & 0.640 &   2.30 & 3.31 & 0.21 & 23.38 &  p &  1.76 &-20.58    \\
  565 &12:36:51.96 &62:13:32.2  & 22.14 & 0.99 &0.917  &0.961  & 0.310 &  -5.00 & 1.15 & 0.17 & 21.33 &  f &  1.34 &-21.02    \\
  575 &12:36:48.97 &62:13:21.9  & 23.84 & 0.71 &1.226  &   -1  & 0.860 &  -5.40 & 3.70 & 0.91 & 25.61 &  p &  3.37 &-15.86    \\
  576 &12:36:49.07 &62:13:21.9  & 22.54 & 1.81 &1.078  &   -1  & 0.060 &  -0.29 & 3.08 & 0.41 & 22.93 &  g &  3.29 &-20.85    \\
  577 &12:36:42.66 &62:13:06.0  & 22.22 & 1.77 &1.285  &   -1  & 0.190 &   3.58 & 1.74 & 1.73 & 24.97 &  p & 14.67 &-22.72    \\
  582 &12:36:46.16 &62:13:13.9  & 23.88 & 0.34 &0.614  &   -1  & 0.620 &  -5.40 & 0.80 & 0.11 & 22.05 &  g &  0.73 &-18.11    \\
  604 &12:36:44.76 &62:13:07.0  & 23.95 & 0.64 &2.776  &   -1  & 0.760 &   6.38 & 3.66 & 1.13 & 27.36 &  p &  3.34 &-14.46    \\
  605 &12:36:44.58 &62:13:04.6  & 20.01 & 1.24 &0.449  &0.485  & 0.050 &   3.29 & 1.68 & 1.17 & 22.64 &  p &  7.03 &-21.28    \\
  607 &12:36:48.78 &62:13:18.5  & 22.36 & 0.41 &0.768  &0.753  & 0.600 &   5.43 & 1.40 & 1.09 & 25.25 &  p &  8.02 &-20.26    \\
  608 &12:36:48.46 &62:13:16.7  & 22.53 & 0.78 &0.502  &0.474  & 0.250 &  -3.14 & 2.41 & 0.37 & 23.18 &  g &  2.35 &-19.18    \\
  609 &12:36:48.26 &62:13:13.8  & 22.11 & 1.80 &1.158  &   -1  & 0.130 &  -3.29 & 1.25 & 0.51 & 23.44 &  g &  4.12 &-21.38    \\
  612 &12:36:44.88 &62:13:04.7  & 23.34 & 1.14 &1.385  &   -1  & 0.540 &  -3.80 & 1.64 & 0.13 & 21.94 &  g &  1.10 &-20.92    \\
  613 &12:36:46.74 &62:13:12.3  & 23.25 & 0.35 &0.599  &   -1  & 0.640 &   2.00 & 2.22 & 0.47 & 24.42 &  f &  3.14 &-18.80    \\
  614 &12:36:45.62 &62:13:08.8  & 23.08 & 0.57 &0.508  &   -1  & 0.430 &  -3.71 & 2.63 & 0.29 & 23.33 &  g &  1.78 &-18.52    \\
  643 &12:36:51.05 &62:13:20.7  & 19.38 & 0.24 &0.193  &0.199  & 0.590 &   4.57 & 1.13 & 1.99 & 23.44 &  p &  6.55 &-20.11    \\
  653 &12:36:56.64 &62:13:39.9  & 23.87 & 1.42 &1.783  &   -1  & 0.420 &   3.50 & 3.61 & 0.41 & 24.28 &  p &  3.43 &-21.68    \\
  659 &12:36:49.46 &62:13:16.7  & 21.72 & 1.50 &1.243  &1.238  & 0.290 &   4.00 & 1.29 & 0.61 & 23.18 &  f &  5.09 &-22.30    \\
  670 &12:36:48.07 &62:13:09.0  & 19.44 & 0.98 &0.553  &0.476  & 0.150 &  -5.71 & 1.89 & 0.63 & 21.61 &  f &  3.74 &-21.86    \\
  680 &12:36:54.73 &62:13:28.0  & 18.57 & 0.68 &0.521  &    0  & 0.960 &  -9.71 & 0.76 & 0.25 & 19.53 &  p &  0.00 &-12.45    \\
  685 &12:36:50.46 &62:13:16.1  & 21.41 & 1.25 &0.909  &0.851  & 0.180 &   1.14 & 2.59 & 1.09 & 23.78 &  p &  8.49 &-21.64    \\
  700 &12:36:49.03 &62:13:09.8  & 23.62 & 0.41 &0.856  &   -1  & 0.680 &  -2.00 & 2.55 & 0.22 & 23.11 &  f &  1.66 &-19.22    \\
  701 &12:36:53.25 &62:13:21.5  & 23.78 & 0.68 &1.982  &   -1  & 0.790 &   1.30 & 2.49 & 0.29 & 23.96 &  f &  2.42 &-21.08    \\
  717 &12:36:49.37 &62:13:11.3  & 21.53 & 0.59 &0.497  &0.477  & 0.440 &  -5.00 & 0.00 & 0.19 & 20.85 &  f &  1.13 &-19.90    \\
  725 &12:36:56.12 &62:13:29.7  & 22.48 & 0.92 &1.151  &1.238  & 0.550 &   2.79 & 1.98 & 0.47 & 23.44 &  p &  1.76 &-17.09    \\
  731 &12:36:44.11 &62:12:44.8  & 21.58 & 2.83 &1.661  &1.755  & 0.060 &  -3.29 & 2.93 & 0.63 & 23.38 &  g &  5.33 &-23.51    \\
  732 &12:36:44.43 &62:12:44.1  & 23.27 & 0.74 &1.982  &   -1  & 0.760 &   7.86 & 3.01 & 1.25 & 26.17 &  p & 10.43 &-21.66    \\
  734 &12:36:43.97 &62:12:50.1  & 20.02 & 1.01 &0.511  &0.557  & 0.190 &   2.64 & 1.52 & 1.61 & 22.80 &  p & 10.40 &-21.71    \\
  735 &12:36:44.19 &62:12:47.8  & 21.34 & 0.34 &0.587  &0.555  & 0.680 &   5.40 & 3.05 & 1.71 & 24.58 &  f & 11.03 &-20.54    \\
  741 &12:36:53.18 &62:13:22.7  & 23.99 & 0.90 &2.636  &2.489  & 0.720 &  -3.30 & 2.00 & 0.23 & 23.63 &  f &  1.86 &-21.69    \\
  748 &12:36:38.60 &62:12:33.8  & 23.76 & 0.34 &0.425  &0.904  & 0.590 &   1.58 & 2.94 & 0.39 & 24.05 &  f &  2.17 &-17.44    \\
  758 &12:36:41.85 &62:12:43.6  & 23.68 & 1.54 &2.111  &   -1  & 0.480 &   4.58 & 2.82 & 0.61 & 25.25 &  p &  5.01 &-22.05    \\
  764 &12:36:49.68 &62:13:07.4  & 23.85 & 0.38 &0.921  &   -1  & 0.690 &  -0.92 & 3.90 & 0.17 & 23.09 &  g &  1.32 &-19.19    \\
  775 &12:36:49.71 &62:13:13.1  & 20.40 & 1.07 &0.452  &0.475  & 0.140 &  -0.29 & 2.74 & 1.07 & 22.51 &  p &  6.36 &-20.83    \\
  779 &12:36:38.41 &62:12:31.3  & 21.47 & 1.26 &0.516  &0.944  & 0.050 &   0.79 & 2.75 & 0.97 & 23.50 &  p &  6.07 &-20.03    \\
  782 &12:36:45.02 &62:12:51.1  & 23.20 & 0.97 &1.761  &2.801  & 0.680 &   4.36 & 1.99 & 1.01 & 25.61 &  p &  7.94 &-22.85    \\
  784 &12:36:56.13 &62:13:25.2  & 22.66 & 2.42 &1.316  &   -1  & 0.050 &  -5.00 & 0.00 & 0.11 & 21.01 &  g &  0.93 &-21.83    \\
  789 &12:36:43.15 &62:12:42.2  & 20.23 & 1.47 &0.838  &0.849  & 0.010 &  -4.71 & 0.76 & 0.31 & 20.66 &  g &  2.37 &-22.70    \\
  800 &12:36:39.98 &62:12:33.6  & 23.62 & 0.91 &1.026  &   -1  & 0.460 &   0.75 & 2.44 & 0.15 & 22.27 &  g &  1.21 &-19.77    \\
  811 &12:36:54.16 &62:13:16.5  & 23.53 & 0.86 &1.202  &   -1  & 0.630 &  -0.50 & 2.40 & 0.13 & 21.80 &  g &  1.08 &-20.22    \\
  813 &12:36:47.72 &62:12:55.8  & 23.64 & 0.98 &3.024  &2.931  & 0.610 &   7.42 & 2.73 & 0.95 & 25.61 &  p &  7.38 &-22.03    \\
  814 &12:36:47.86 &62:12:55.4  & 23.65 & 1.27 &2.944  &2.931  & 0.460 &   8.20 & 2.17 & 1.33 & 26.17 &  p & 10.33 &-22.60    \\
  819 &12:36:55.02 &62:13:18.8  & 23.75 & 1.41 &0.849  &   -1  & 0.060 &  -4.00 & 1.26 & 0.17 & 22.93 &  g &  1.30 &-19.10    \\
  822 &12:36:39.87 &62:12:31.5  & 23.79 & 0.42 &0.717  &   -1  & 0.460 &   1.08 & 2.75 & 0.17 & 22.59 &  g &  1.22 &-18.65    \\
  827 &12:36:39.78 &62:12:28.5  & 23.68 & 0.14 &0.052  &   -1  & 0.800 &   6.10 & 3.72 & 1.33 & 26.17 &  p &  3.46 &-15.28    \\
  829 &12:36:40.88 &62:12:34.0  & 23.46 & 0.73 &4.264  &   -1  & 0.520 &  -6.60 & 2.10 & 0.05 & 20.38 &  p &  0.00 &-23.26    \\
  833 &12:36:45.98 &62:12:50.4  & 23.96 & 0.45 &3.809  &   -1  & 0.650 &  -6.60 & 2.10 & 0.05 & 20.58 &  p &  0.00 &-24.37    \\
  843 &12:36:54.71 &62:13:14.8  & 23.56 & 0.78 &2.255  &2.232  & 0.750 &  -2.30 & 3.90 & 0.17 & 22.64 &  g &  1.40 &-21.71    \\
  846 &12:36:47.54 &62:12:52.7  & 23.60 & 0.41 &0.714  &0.681  & 0.580 &   2.25 & 2.36 & 0.43 & 24.38 &  f &  3.04 &-18.75    \\
  847 &12:36:46.13 &62:12:46.5  & 21.16 & 1.20 &0.765  &  0.9  & 0.070 &  -4.43 & 2.51 & 0.39 & 22.19 &  g &  3.04 &-21.94    \\
  848 &12:36:54.99 &62:13:14.8  & 23.71 & 0.18 &0.536  &0.511  & 0.640 &  -0.60 & 1.74 & 0.29 & 23.50 &  f &  1.79 &-17.98    \\
  855 &12:36:49.63 &62:12:57.6  & 20.92 & 1.09 &0.456  &0.475  & 0.160 &  -0.93 & 3.37 & 0.19 & 19.81 &  f &  1.13 &-20.35    \\
  860 &12:36:44.18 &62:12:40.4  & 22.98 & 0.35 &0.888  &0.875  & 0.660 &   6.71 & 1.89 & 1.73 & 26.17 &  p & 13.37 &-19.96    \\
  864 &12:36:41.04 &62:12:30.4  & 23.51 & 1.28 &1.508  &   -1  & 0.490 &   6.50 & 1.66 & 0.63 & 25.25 &  f &  5.34 &-21.16    \\
  868 &12:36:53.65 &62:13:08.3  & 20.84 & 0.42 &0.108  &    0  & 0.690 &  -9.29 & 1.89 & 0.11 & 19.67 &  p &  0.00 &-10.43    \\
  870 &12:36:51.40 &62:13:00.6  & 23.10 & 0.08 &0.055  & 0.09  & 0.740 &   9.71 & 0.49 & 1.99 & 27.36 &  p &  3.31 &-14.49    \\
  880 &12:36:48.32 &62:12:49.8  & 19.83 & 1.11 &0.726  &    0  & 1.000 & -10.00 & 0.00 & 0.15 & 19.42 &  p &  0.00 &-10.76    \\
  882 &12:36:45.66 &62:12:41.9  & 22.27 & 3.28 &1.792  &   -1  & 0.040 &  -3.29 & 2.98 & 0.21 & 21.86 &  f &  1.77 &-23.03    \\
  884 &12:36:55.16 &62:13:09.0  & 23.61 & 1.49 &1.269  &   -1  & 0.300 &   2.12 & 3.79 & 0.53 & 24.76 &  p &  4.46 &-20.75    \\
  885 &12:36:55.14 &62:13:11.4  & 23.38 & 0.32 &0.228  &0.321  & 0.710 &   2.80 & 3.25 & 0.71 & 24.97 &  g &  3.85 &-17.76    \\
  886 &12:36:55.45 &62:13:11.2  & 20.44 & 1.75 &0.973  &0.968  & 0.020 &  -4.57 & 1.13 & 0.37 & 21.30 &  g &  2.94 &-22.86    \\
  893 &12:36:38.98 &62:12:19.7  & 21.68 & 0.57 &0.623  &0.609  & 0.450 &   3.50 & 3.71 & 1.57 & 24.76 &  p & 10.58 &-20.38    \\
  897 &12:36:50.81 &62:12:55.9  & 22.34 & 0.26 &0.311  &0.321  & 0.690 &   0.93 & 2.81 & 0.37 & 22.67 &  f &  1.72 &-18.26    \\
  909 &12:36:57.73 &62:13:15.2  & 22.24 & 0.64 &0.843  &0.952  & 0.550 &   8.21 & 2.51 & 1.11 & 24.97 &  p &  8.79 &-20.98    \\
  913 &12:36:45.00 &62:12:39.6  & 23.20 & 0.95 &1.249  &1.225  & 0.600 &   5.36 & 1.60 & 1.39 & 26.17 &  p & 11.57 &-20.70    \\
  921 &12:36:49.46 &62:12:48.8  & 23.13 & 0.83 &0.678  &   -1  & 0.270 &  -3.50 & 2.60 & 0.17 & 22.21 &  g &  1.20 &-19.19    \\
  939 &12:36:50.83 &62:12:51.5  & 22.82 & 0.38 &0.526  &0.485  & 0.560 &   6.79 & 2.31 & 1.35 & 26.17 &  p &  8.40 &-18.84    \\
  942 &12:36:48.99 &62:12:45.9  & 23.50 & 0.29 &0.484  &0.512  & 0.580 &   3.80 & 5.40 & 0.71 & 24.97 &  f &  4.23 &-18.09    \\
  945 &12:36:58.28 &62:13:14.9  & 23.41 & 2.07 &1.136  &   -1  & 0.050 &  -4.60 & 0.80 & 0.21 & 22.93 &  f &  1.73 &-20.28    \\
  953 &12:36:54.18 &62:13:01.1  & 23.78 & 1.19 &1.197  &   -1  & 0.400 &   5.25 & 3.71 & 0.33 & 23.96 &  p &  2.75 &-20.09    \\
  954 &12:36:46.77 &62:12:37.1  & 21.61 & 0.78 &0.708  &    0  & 0.980 &  -8.33 & 2.58 & 0.09 & 19.83 &  p &  0.00 & -9.28    \\
  955 &12:36:47.04 &62:12:36.9  & 20.59 & 0.47 &0.432  &0.321  & 0.530 &  -2.00 & 2.00 & 0.65 & 22.49 &  g &  3.04 &-19.94    \\
  956 &12:36:55.15 &62:13:03.6  & 21.97 & 1.60 &0.905  &0.952  & 0.030 &  -4.71 & 0.76 & 0.11 & 20.14 &  g &  0.85 &-21.07    \\
  971 &12:36:50.26 &62:12:45.8  & 20.19 & 1.22 &0.753  & 0.68  & 0.080 &  -3.14 & 1.34 & 0.53 & 21.67 &  g &  3.75 &-22.09    \\
  972 &12:36:39.72 &62:12:14.1  & 22.31 & 1.61 &0.989  &   -1  & 0.090 &   4.21 & 0.76 & 0.71 & 23.70 &  f &  5.65 &-20.96    \\
  975 &12:36:57.21 &62:13:07.7  & 22.55 & 2.03 &1.102  &   -1  & 0.040 &  -5.83 & 2.04 & 0.09 & 20.37 &  g &  0.74 &-21.19    \\
  983 &12:36:39.43 &62:12:11.7  & 22.29 & 1.58 &0.974  &   -1  & 0.080 &  -3.80 & 2.40 & 0.11 & 20.57 &  g &  0.87 &-20.95    \\
  989 &12:36:44.64 &62:12:27.4  & 22.61 & 1.16 &1.444  &  2.5  & 0.520 &   0.30 & 4.06 & 0.31 & 22.83 &  f &  2.50 &-23.02    \\
  994 &12:36:53.89 &62:12:54.1  & 19.83 & 1.08 &0.713  &0.642  & 0.190 &   2.86 & 1.57 & 1.03 & 22.49 &  p &  7.11 &-22.28    \\
 1010 &12:36:56.92 &62:13:01.7  & 21.07 & 1.81 &1.196  &   -1  & 0.140 &  -4.57 & 1.13 & 0.49 & 22.56 &  g &  4.11 &-23.09    \\
 1011 &12:36:56.98 &62:12:56.5  & 23.68 & 0.89 & 1.08  &   -1  & 0.520 &   4.12 & 1.51 & 0.71 & 25.61 &  f &  5.83 &-19.88    \\
 1014 &12:36:41.49 &62:12:15.0  & 21.63 & 1.83 &1.024  &   -1  & 0.040 &  -2.33 & 2.29 & 0.11 & 19.74 &  f &  0.88 &-21.76    \\
 1021 &12:36:46.21 &62:12:28.5  & 23.74 & 1.08 &1.453  &   -1  & 0.570 &  -0.40 & 7.90 & 0.61 & 25.25 &  p &  5.16 &-20.54    \\
 1022 &12:36:50.22 &62:12:39.8  & 20.17 & 0.52 &0.478  &0.474  & 0.460 &   6.10 & 1.85 & 1.99 & 23.70 &  p & 11.81 &-21.24    \\
 1023 &12:36:47.78 &62:12:32.9  & 22.40 & 0.96 &0.941  & 0.96  & 0.350 &  -0.21 & 4.85 & 0.23 & 22.02 &  p &  1.82 &-20.85    \\
 1024 &12:36:56.93 &62:12:58.3  & 23.01 & 0.66 &0.499  & 0.52  & 0.310 &   0.30 & 2.44 & 0.47 & 23.70 &  f &  2.93 &-18.61    \\
 1027 &12:36:42.92 &62:12:16.3  & 20.11 & 0.64 & 0.45  &0.454  & 0.390 &   3.36 & 1.80 & 1.99 & 23.63 &  p & 11.53 &-21.18    \\
 1029 &12:36:47.28 &62:12:30.7  & 22.45 & 0.36 &0.446  &0.421  & 0.580 &   3.29 & 2.67 & 0.51 & 23.63 &  f &  2.83 &-18.74    \\
 1031 &12:36:40.01 &62:12:07.3  & 20.91 & 1.39 &0.887  &1.015  & 0.100 &  -4.00 & 2.24 & 0.29 & 21.27 &  f &  2.33 &-22.51    \\
 1036 &12:36:43.63 &62:12:18.3  & 22.22 & 0.53 &0.761  &0.752  & 0.490 &  -0.86 & 3.81 & 0.35 & 22.80 &  f &  2.57 &-20.41    \\
 1042 &12:36:57.30 &62:12:59.7  & 21.07 & 0.29 &0.492  &0.475  & 0.640 &   3.90 & 2.09 & 1.91 & 24.76 &  f & 11.34 &-20.37    \\
 1047 &12:36:49.56 &62:12:36.0  & 23.48 & 1.18 &1.993  &   -1  & 0.620 &   1.20 & 0.45 & 0.21 & 22.76 &  f &  1.75 &-21.76    \\
 1050 &12:36:56.57 &62:12:57.4  & 23.31 & 2.31 &1.197  &   -1  & 0.030 &  -5.00 & 1.90 & 0.05 & 19.95 &  f &  0.42 &-20.90    \\
 1051 &12:36:58.06 &62:13:00.4  & 22.12 & 0.27 &0.321  & 0.32  & 0.710 &   0.50 & 2.90 & 0.31 & 22.25 &  f &  1.45 &-18.45    \\
 1076 &12:36:40.84 &62:12:03.1  & 22.61 & 0.61 &0.984  & 1.01  & 0.640 &   9.60 & 0.49 & 0.79 & 24.42 &  f &  6.35 &-20.73    \\
 1077 &12:36:40.75 &62:12:04.9  & 23.56 & 1.06 &0.919  &   -1  & 0.260 &  -0.50 & 3.50 & 0.17 & 22.67 &  g &  1.33 &-19.57    \\
 1078 &12:36:40.96 &62:12:05.3  & 22.52 & 0.78 &2.666  &0.882  & 0.680 &  -6.25 & 2.16 & 0.07 & 19.98 &  f &  0.56 &-22.35    \\
 1080 &12:36:54.04 &62:12:45.6  & 22.00 & 0.74 &0.762  &    0  & 0.990 &  -5.83 & 3.44 & 0.07 & 19.52 &  p &  0.00 & -8.96    \\
 1086 &12:36:55.61 &62:12:49.2  & 22.67 & 0.41 &0.903  & 0.95  & 0.650 &   7.08 & 2.84 & 1.63 & 26.17 &  p & 12.90 &-20.46    \\
 1087 &12:36:55.24 &62:12:52.5  & 23.94 & 0.33 &0.699  &   -1  & 0.730 &   8.36 & 2.84 & 0.83 & 26.17 &  p &  5.79 &-18.28    \\
 1090 &12:36:56.60 &62:12:52.7  & 23.52 & 0.82 & 1.27  &1.231  & 0.740 &   2.40 & 6.10 & 1.99 & 26.17 &  p & 16.59 &-20.27    \\
 1091 &12:36:56.72 &62:12:52.6  & 21.77 & 1.98 &1.219  &   -1  & 0.110 &  -3.86 & 2.27 & 0.31 & 22.21 &  f &  2.58 &-22.24    \\
 1092 &12:36:44.56 &62:12:15.5  & 23.06 & 2.32 &1.751  &   -1  & 0.140 &   5.00 & 4.55 & 0.69 & 24.76 &  p &  5.81 &-22.45    \\
 1115 &12:36:41.26 &62:12:03.0  & 23.89 & 0.49 &3.226  &3.216  & 0.750 &   8.75 & 2.60 & 1.03 & 26.17 &  p &  7.77 &-23.58    \\
 1117 &12:36:41.95 &62:12:05.4  & 20.22 & 0.78 &0.459  &0.432  & 0.310 &  -1.00 & 2.77 & 0.61 & 21.78 &  p &  3.44 &-20.90    \\
 1120 &12:36:55.56 &62:12:45.5  & 21.08 & 0.90 &0.785  & 0.79  & 0.300 &  -3.86 & 2.04 & 0.33 & 21.73 &  f &  2.47 &-21.65    \\
 1127 &12:36:56.64 &62:12:45.5  & 18.94 & 1.05 &0.589  &0.518  & 0.150 &   2.57 & 1.62 & 1.99 & 22.67 &  p & 12.40 &-22.57    \\
 1128 &12:36:45.42 &62:12:13.6  & 22.12 &-0.98 &0.000  &    0  & 0.950 &  -8.00 & 3.46 & 0.07 & 19.89 &  p &  0.00 &-10.57    \\
 1135 &12:36:59.30 &62:12:55.8  & 21.53 & 0.95 &0.764  &    0  & 1.000 &  -7.83 & 2.27 & 0.07 & 19.42 &  p &  0.00 & -9.21    \\
 1136 &12:36:58.76 &62:12:52.4  & 20.89 & 0.40 &0.422  &0.321  & 0.540 &   3.86 & 1.22 & 1.99 & 24.76 &  f &  9.29 &-19.61    \\
 1141 &12:36:49.98 &62:12:26.3  & 23.69 & 0.86 &1.234  &   -1  & 0.650 &   2.10 & 4.56 & 0.55 & 24.58 &  p &  4.53 &-20.64    \\
 1160 &12:36:50.84 &62:12:27.2  & 23.82 & 0.46 &0.707  &   -1  & 0.500 &   5.58 & 3.32 & 0.47 & 24.97 &  f &  3.36 &-18.58    \\
 1166 &12:36:49.95 &62:12:25.5  & 23.26 & 1.10 &1.234  &1.205  & 0.490 &  -5.00 & 0.00 & 0.09 & 21.16 &  g &  0.35 &-16.41    \\
 1168 &12:36:49.06 &62:12:21.2  & 21.86 & 0.62 &0.915  &0.953  & 0.550 &   5.10 & 3.29 & 0.45 & 23.13 &  p &  3.56 &-21.37    \\
 1171 &12:36:53.44 &62:12:34.3  & 22.18 & 0.67 &0.576  & 0.56  & 0.410 &   1.57 & 2.28 & 0.43 & 22.73 &  p &  2.79 &-19.60    \\
 1198 &12:36:49.53 &62:12:20.1  & 23.91 & 0.56 & 0.94  &0.961  & 0.620 &   1.17 & 2.21 & 0.41 & 24.28 &  p &  3.25 &-19.35    \\
 1208 &12:36:52.86 &62:12:29.6  & 23.73 & 0.89 &0.701  &   -1  & 0.220 &   2.43 & 1.24 & 0.47 & 24.76 &  g &  3.37 &-18.64    \\
 1211 &12:36:48.63 &62:12:15.8  & 22.16 & 2.38 &1.711  &   -1  & 0.120 &  -1.29 & 2.98 & 0.29 & 22.35 &  g &  2.44 &-23.33    \\
 1213 &12:36:48.25 &62:12:13.8  & 21.75 & 0.85 &0.949  &0.962  & 0.430 &   4.57 & 1.13 & 1.99 & 25.61 &  p & 15.80 &-21.45    \\
 1214 &12:36:52.09 &62:12:26.3  & 21.83 & 2.03 &1.119  &   -1  & 0.040 &  -3.80 & 2.40 & 0.26 & 22.00 &  g &  2.11 &-21.71    \\
 1229 &12:37:00.11 &62:12:49.9  & 23.37 & 1.07 &1.291  &   -1  & 0.560 &  -1.50 & 3.32 & 0.09 & 21.12 &  g &  0.75 &-20.46    \\
 1231 &12:36:55.03 &62:12:34.2  & 22.38 & 1.87 &1.108  &   -1  & 0.080 &   0.25 & 3.03 & 0.11 & 20.29 &  g &  0.91 &-21.40    \\
 1240 &12:36:50.16 &62:12:17.0  & 20.96 & 1.48 &0.908  &0.905  & 0.070 &   2.50 & 2.71 & 0.49 & 21.99 &  f &  3.80 &-22.08    \\
 1247 &12:36:56.35 &62:12:41.1  & 19.02 & 0.25 &0.000  &    0  & 0.340 &  -9.29 & 1.89 & 0.21 & 19.42 &  p &  0.00 &-12.43    \\
 1253 &12:36:51.71 &62:12:20.2  & 20.69 & 0.89 & 0.43  &0.401  & 0.230 &  -0.64 & 2.69 & 0.47 & 21.43 &  f &  2.53 &-20.19    \\
 1258 &12:36:44.83 &62:12:00.2  & 22.62 & 0.43 &0.487  &0.457  & 0.530 &   3.58 & 2.05 & 1.35 & 25.61 &  f &  7.85 &-18.74    \\
 1266 &12:36:46.41 &62:12:04.6  & 23.94 & 0.70 &1.175  &   -1  & 0.810 &   9.33 & 1.21 & 1.95 & 27.36 &  p & 14.84 &-18.68    \\
 1282 &12:36:46.95 &62:12:05.3  & 23.87 & 0.24 &0.051  &   -1  & 0.640 &   4.37 & 1.85 & 0.35 & 24.42 &  f &  1.95 &-17.35    \\
 1286 &12:36:45.96 &62:12:01.4  & 23.29 & 0.48 &0.733  &0.679  & 0.540 &   1.00 & 2.36 & 0.43 & 23.96 &  f &  3.04 &-19.10    \\
 1305 &12:36:53.42 &62:12:21.7  & 23.69 & 0.35 &1.877  &   -1  & 0.850 &   5.25 & 2.95 & 0.35 & 24.28 &  f &  2.93 &-21.49    \\
 1315 &12:36:51.61 &62:12:17.3  & 23.81 & 1.24 &1.945  &   -1  & 0.510 &  -2.00 & 2.55 & 0.07 & 20.90 &  g &  0.58 &-21.62    \\
 1316 &12:36:43.42 &62:11:51.5  & 22.37 & 0.67 &1.191  &1.242  & 0.750 &  -3.33 & 2.62 & 0.13 & 20.82 &  g &  1.08 &-21.54    \\
 1325 &12:36:52.67 &62:12:19.7  & 22.81 & 0.42 & 0.46  &0.401  & 0.550 &  -3.20 & 1.47 & 0.15 & 21.37 &  f &  0.81 &-18.27    \\
 1335 &12:36:41.41 &62:11:42.5  & 21.04 & 1.93 &1.499  &1.524  & 0.220 &   4.50 & 1.32 & 1.93 & 24.58 &  p & 16.32 &-24.13    \\
 1336 &12:36:41.32 &62:11:40.8  & 21.47 & 1.32 &0.651  &0.585  & 0.050 &  -2.64 & 2.87 & 0.45 & 22.29 &  f &  2.98 &-20.35    \\
 1348 &12:36:43.19 &62:11:48.0  & 20.75 & 1.69 &1.022  & 1.01  & 0.080 &  -0.29 & 2.75 & 0.71 & 22.76 &  p &  5.70 &-22.66    \\
 1354 &12:36:45.06 &62:11:54.1  & 23.55 & 0.65 &0.879  &   -1  & 0.930 &   2.38 & 5.76 & 1.07 & 26.17 &  f &  8.27 &-19.36    \\
 1355 &12:36:45.66 &62:11:53.9  & 23.07 & 1.35 &1.157  &   -1  & 0.320 &  -1.00 & 2.12 & 0.19 & 22.42 &  g &  1.57 &-20.66    \\
 1356 &12:36:45.33 &62:11:54.5  & 21.33 & 1.78 &1.015  &   -1  & 0.050 &  -3.67 & 1.97 & 0.31 & 21.77 &  g &  2.48 &-22.03    \\
 1357 &12:36:45.41 &62:11:53.1  & 23.21 & 0.93 &3.028  &2.803  & 0.620 &  -0.90 & 3.54 & 0.99 & 25.25 &  f &  7.78 &-22.81    \\
 1358 &12:36:45.30 &62:11:52.2  & 23.49 & 0.63 &3.017  &2.803  & 0.740 &   7.92 & 2.91 & 0.97 & 25.25 &  f &  7.62 &-22.83    \\
 1359 &12:36:49.31 &62:12:07.3  & 23.76 & 1.07 &0.921  &   -1  & 0.250 &   0.60 & 2.48 & 0.19 & 22.80 &  f &  1.49 &-19.36    \\
 1364 &12:36:42.15 &62:11:44.7  & 23.82 & 0.72 &1.506  &   -1  & 0.790 &   6.50 & 4.05 & 0.33 & 24.28 &  f &  2.79 &-19.93    \\
 1411 &12:36:52.01 &62:12:09.7  & 22.59 & 0.33 &0.506  &0.458  & 0.600 &  -0.83 & 3.29 & 0.83 & 24.97 &  g &  4.83 &-18.80    \\
 1414 &12:36:57.20 &62:12:25.8  & 22.03 & 0.42 &0.616  &0.561  & 0.600 &   2.75 & 1.37 & 0.71 & 24.28 &  p &  4.60 &-19.83    \\
 1418 &12:36:43.81 &62:11:42.8  & 19.56 & 1.29 &0.763  &0.765  & 0.030 &  -5.00 & 0.00 & 0.61 & 21.46 &  g &  4.51 &-23.07    \\
 1429 &12:37:00.56 &62:12:34.7  & 20.42 & 0.98 &0.577  &0.563  & 0.130 &  -5.71 & 1.89 & 0.33 & 21.10 &  g &  2.14 &-21.39    \\
 1434 &12:36:46.52 &62:11:51.3  & 20.97 & 1.02 &0.546  &0.503  & 0.130 &  -4.71 & 0.76 & 0.11 & 19.42 &  g &  0.68 &-20.53    \\
 1436 &12:36:54.78 &62:12:16.6  & 23.58 & 0.80 &1.742  &   -1  & 0.740 &   3.42 & 0.97 & 0.45 & 24.16 &  f &  3.77 &-20.94    \\
 1446 &12:36:44.49 &62:11:43.7  & 23.79 & 0.47 & 0.98  & 1.02  & 0.730 &  10.00 & 0.00 & 1.13 & 26.17 &  p &  9.10 &-19.55    \\
 1447 &12:36:44.46 &62:11:41.6  & 22.52 & 0.62 &0.967  & 1.02  & 0.610 &   5.86 & 2.46 & 1.91 & 26.17 &  p & 15.38 &-20.87    \\
 1453 &12:36:56.64 &62:12:20.1  & 20.91 & 1.57 &0.933  & 0.93  & 0.060 &  -3.00 & 2.52 & 0.37 & 21.80 &  f &  2.90 &-22.21    \\
 1462 &12:36:41.63 &62:11:31.8  & 19.64 & 0.18 &0.083  &0.089  & 0.740 &   2.00 & 1.55 & 1.99 & 23.44 &  p &  3.31 &-18.13    \\
 1469 &12:36:42.30 &62:11:34.7  & 23.16 & 0.81 &0.753  &   -1  & 0.330 &   2.50 & 2.10 & 0.61 & 24.76 &  f &  4.53 &-19.40    \\
 1474 &12:36:47.45 &62:11:50.8  & 23.49 & 0.59 &0.884  &   -1  & 0.500 &   0.83 & 2.54 & 0.13 & 21.67 &  f &  1.01 &-19.55    \\
 1481 &12:36:45.31 &62:11:42.8  & 23.78 & 0.20 &0.66   &0.558  & 0.700 &   6.83 & 4.17 & 0.33 & 23.87 &  p &  2.13 &-18.08    \\
 1482 &12:36:49.35 &62:11:55.0  & 23.24 & 0.24 &0.998  &0.961  & 0.800 &  -3.20 & 1.83 & 0.05 & 19.53 &  f &  0.40 &-20.02    \\
 1486 &12:36:55.37 &62:12:13.4  & 23.71 & 0.97 &4.611  &   -1  & 0.410 &  -6.60 & 2.10 & 0.13 & 22.39 &  p &  0.85 &-23.14    \\
 1488 &12:36:46.18 &62:11:42.1  & 19.88 & 1.42 &1.088  &1.013  & 0.230 &   3.71 & 1.11 & 1.97 & 23.70 &  p & 15.85 &-23.55    \\
 1495 &12:36:46.87 &62:11:44.9  & 22.57 & 0.74 &1.072  & 1.06  & 0.660 &   4.00 & 1.53 & 1.79 & 26.17 &  p & 14.54 &-20.83    \\
 1510 &12:36:58.64 &62:12:21.7  & 23.00 & 0.42 &0.71   &0.682  & 0.610 &  -4.80 & 3.03 & 0.15 & 21.75 &  f &  1.06 &-19.38    \\
 1512 &12:36:52.10 &62:12:01.2  & 23.48 & 2.76 &1.534  &   -1  & 0.060 &   6.75 & 3.46 & 0.77 & 25.61 &  f &  6.53 &-21.22    \\
 1513 &12:37:00.07 &62:12:25.3  & 23.06 & 0.77 &2.379  & 2.05  & 0.740 &   6.30 & 2.96 & 0.41 & 23.78 &  p &  3.35 &-21.80    \\
 1521 &12:36:52.56 &62:12:01.6  & 23.24 & 0.30 &0.465  &   -1  & 0.480 &  -1.00 & 6.16 & 0.19 & 22.70 &  f &  1.14 &-18.34    \\
 1522 &12:36:43.90 &62:11:34.1  & 22.71 & 1.83 &1.028  &   -1  & 0.050 &  -5.00 & 0.00 & 0.05 & 19.56 &  g &  0.40 &-20.65    \\
 1523 &12:36:44.39 &62:11:33.1  & 19.34 & 1.99 &1.038  & 1.05  & 0.010 &  -5.00 & 0.00 & 0.85 & 21.87 &  g &  6.83 &-24.10    \\
 1525 &12:36:49.25 &62:11:48.5  & 21.61 & 1.37 &0.975  &0.961  & 0.170 &  -3.67 & 2.21 & 0.37 & 22.08 &  f &  2.94 &-21.65    \\
 1538 &12:36:55.95 &62:12:10.7  & 23.41 & 1.16 &1.312  &   -1  & 0.470 &  -4.50 & 1.00 & 0.23 & 23.13 &  g &  1.94 &-20.76    \\
 1550 &12:36:58.29 &62:12:16.5  & 23.75 & 1.50 &2.14   &   -1  & 0.470 &   0.60 & 4.30 & 0.15 & 22.46 &  f &  1.23 &-21.99    \\
 1553 &12:37:01.65 &62:12:25.9  & 23.57 & 0.27 &0.929  &0.973  & 0.780 &   6.80 & 3.70 & 0.41 & 24.28 &  f &  3.27 &-19.48    \\
 1558 &12:36:44.75 &62:11:33.4  & 22.65 & 1.80 &0.913  &   -1  & 0.000 &  -6.00 & 2.24 & 0.07 & 20.39 &  f &  0.55 &-20.59    \\
 1562 &12:36:57.58 &62:12:12.7  & 23.44 & 0.33 &0.796  &0.561  & 0.520 &   6.80 & 3.27 & 1.57 & 26.17 &  p & 11.83 &-19.38    \\
 1563 &12:36:57.48 &62:12:10.6  & 19.77 & 1.29 &0.858  &0.665  & 0.070 &  -3.43 & 2.30 & 0.41 & 20.79 &  g &  2.87 &-22.44    \\
 1564 &12:36:51.96 &62:11:55.4  & 23.52 & 0.27 &0.426  &   -1  & 0.530 &   5.40 & 2.30 & 1.05 & 26.17 &  p &  5.93 &-17.55    \\
 1568 &12:36:54.38 &62:12:02.6  & 23.33 & 1.39 &1.233  &   -1  & 0.320 &   6.80 & 3.27 & 0.15 & 21.90 &  f &  1.26 &-20.72    \\
 1569 &12:36:56.42 &62:12:09.3  & 23.08 & 0.29 &0.398  &0.321  & 0.620 &   3.67 & 2.48 & 0.93 & 24.97 &  p &  4.34 &-17.60    \\
 1578 &12:36:58.30 &62:12:14.2  & 22.96 & 0.40 &0.987  &1.020  & 0.710 &  -1.50 & 3.54 & 0.17 & 22.08 &  g &  1.37 &-20.40    \\
 1582 &12:36:49.12 &62:11:50.6  & 23.85 & 1.52 &1.504  &   -1  & 0.370 &   0.12 & 3.71 & 0.23 & 23.23 &  g &  1.95 &-21.03    \\
 1594 &12:37:01.71 &62:12:23.6  & 23.79 & 0.48 &1.114  &1.191  & 0.760 &  -5.40 & 3.70 & 0.19 & 22.86 &  f &  1.55 &-19.69    \\

\enddata

\tablenotetext{a}{J2000}
\tablenotetext{b}{Photometric redshifts}
\tablenotetext{c}{Spectroscopic redshift where available from the
literature--if a star then equal to 0; 
set to -1 if unavailable.}
\tablenotetext{d}{Spectral type, calculated from best-fit spectrum in
photometric redshift procedure, as defined in Budavari et al.\ (2000)}
\tablenotetext{e}{Morphological T-type}
\tablenotetext{f}{Uncertainty in TT}
\tablenotetext{g}{Arcsec, as measured in the \H160 data}
\tablenotetext{h}{mag arcsec$^{-2}$}
\tablenotetext{i}{Quality of the de Vaucouleur fit}
\tablenotetext{j}{The fitted half-light radius in kpc}

\label{samp}
\end{deluxetable}

\end{document}